\documentclass[aps,prl,reprint,superscriptaddress,floatfix]{revtex4-2}
\usepackage[T1]{fontenc}
\usepackage[utf8]{inputenc}
\usepackage[english]{babel}
\usepackage{amsmath,amssymb}
\usepackage{graphicx}
\usepackage{hyperref}
\begin{document}

\title{Atomic Edge Resonators Flatten Phonon Bands in Biphenylene Nanoribbons}
\author{V.~L.~Katkov}

\affiliation{Bogoliubov Laboratory of Theoretical Physics, Joint Institute for Nuclear Research, Dubna, Russia}
\author{N.~M.~Chtchelkatchev}
\affiliation{Bogoliubov Laboratory of Theoretical Physics, Joint Institute for Nuclear Research, Dubna, Russia}
\author{V.~A.~Osipov}
\affiliation{Bogoliubov Laboratory of Theoretical Physics, Joint Institute for Nuclear Research, Dubna, Russia}
\date{\today}
\begin{abstract}
We show that H--F edge complexes in biphenylene nanoribbons act as intrinsic atomic resonators, flattening low-frequency phonon branches. With increasing wave vector, flexural and twisting vibrations concentrate at the edges while motion of the carbon backbone is suppressed. Displacement profiles and edge-mass tuning identify a local-resonance mechanism, an atomic counterpart of a tuned mass damper. Combining machine-learned interatomic potentials, frozen-phonon calculations, and density-functional tight binding, we find that weakly dispersive modes contribute substantially to thermal band-gap narrowing. Zero-point motion shifts the calculated gap closure toward the experimentally observed ribbon widths. A hexagonal boron nitride substrate preserves the principal spectral features while reducing thermal renormalization. These results establish edge chemistry as a route to nearly flat phonon bands and connect a local atomic resonance to the electronic response of an entire nanoribbon.
\end{abstract}
\maketitle
%TC:endignore

Flat bands concentrate states within a narrow energy window, making interactions and the spatial structure of excitations central to material behavior. Recent imaging of interaction-reshaped bands in magic-angle graphene~\cite{Xiao2026FlatBands}, flat-band engineering in a covalent kagome lattice~\cite{Yan2026Covalent}, and the synthesis of graphene nanoribbons with topological flat bands~\cite{Yin2025Flatbands} connect this physics to atomic structure and chemical design. Its vibrational counterpart is advancing rapidly: all-flat phononic bands have been observed in a mechanical metamaterial~\cite{Samak2024AllFlat}, while tunable moir\'e phonons~\cite{Ramos2025Phonons} and localized or directional vibrational energy transport~\cite{Li2026Phononic} have been investigated in layered atomic systems. Can the chemistry of an atomic edge control collective vibrations and their electronic consequences? 
 
Local resonators provide this capability.
Their hybridization with propagating vibrations can flatten phonon branches and suppress thermal transport in nanophononic structures~\cite{Davis2014,Xiong2016,Ma2018,Cui2022,Beardo2024}.  At the atomic scale, however, a flat phonon branch alone does not reveal its microscopic origin or its effect on the electronic gap; these require analysis of the mode eigenvectors and gap-response weights. Recent mode-resolved measurements of electron--phonon coupling in twisted bilayer graphene~\cite{Birkbeck2025Phonons} and many-electron calculations of temperature-dependent semiconductor gaps~\cite{Gong2026Gap} exemplify how lattice vibrations influence electronic structure.

A chemically terminated ribbon represents a distinctive example of this problem. 
The resonators are part of its boundary, so their coupling to the host is set by atomic masses and chemical bonds. The carbon backbone supplies dispersive collective modes, while the edge supplies a softer local coordinate. If these frequency scales overlap, changing the character of a propagating branch need not require a separate patterned resonator structure. Demonstrating this mechanism requires tracking the eigenvectors of the same phonon branch as $q$ varies; weak dispersion alone does not prove edge localization.

Biphenylene provides a chemically accessible platform. Its network of four-, six-, and eight-membered carbon rings has been synthesized on a surface~\cite{Fan2021}. The biphenylene ribbons formed during this interpolymer dehydrofluorination (HF-zipping) process have the alternating H/F edge termination considered here. A new route reported in 2026 produces unsubstituted biphenylene ribbons up to approximately $40$ nm long~\cite{Fan2026Synthesis}. Here we consider H/F-terminated ribbons, both freestanding and supported on monolayer hexagonal boron nitride (hBN). Their width-dependent electronic gap~\cite{Fan2021,Zhou2024,Katkov2026} provides a sensitive probe of vibrational effects, including zero-point renormalization~\cite{Giustino2010,Miglio2020,Engel2022}.

We show that the H--F edge complexes act as intrinsic resonators. As the wave vector increases, twisting and flexural modes evolve from motion distributed across the ribbon into nearly degenerate, weakly dispersive edge vibrations. Their contribution to gap narrowing links local edge dynamics to the electronic spectrum. The hBN-supported system retains the principal spectral signatures, extending the mechanism to a ribbon coupled to a two-dimensional environment.

\begin{figure}[!htbp]
\centering
\includegraphics[width=\columnwidth]{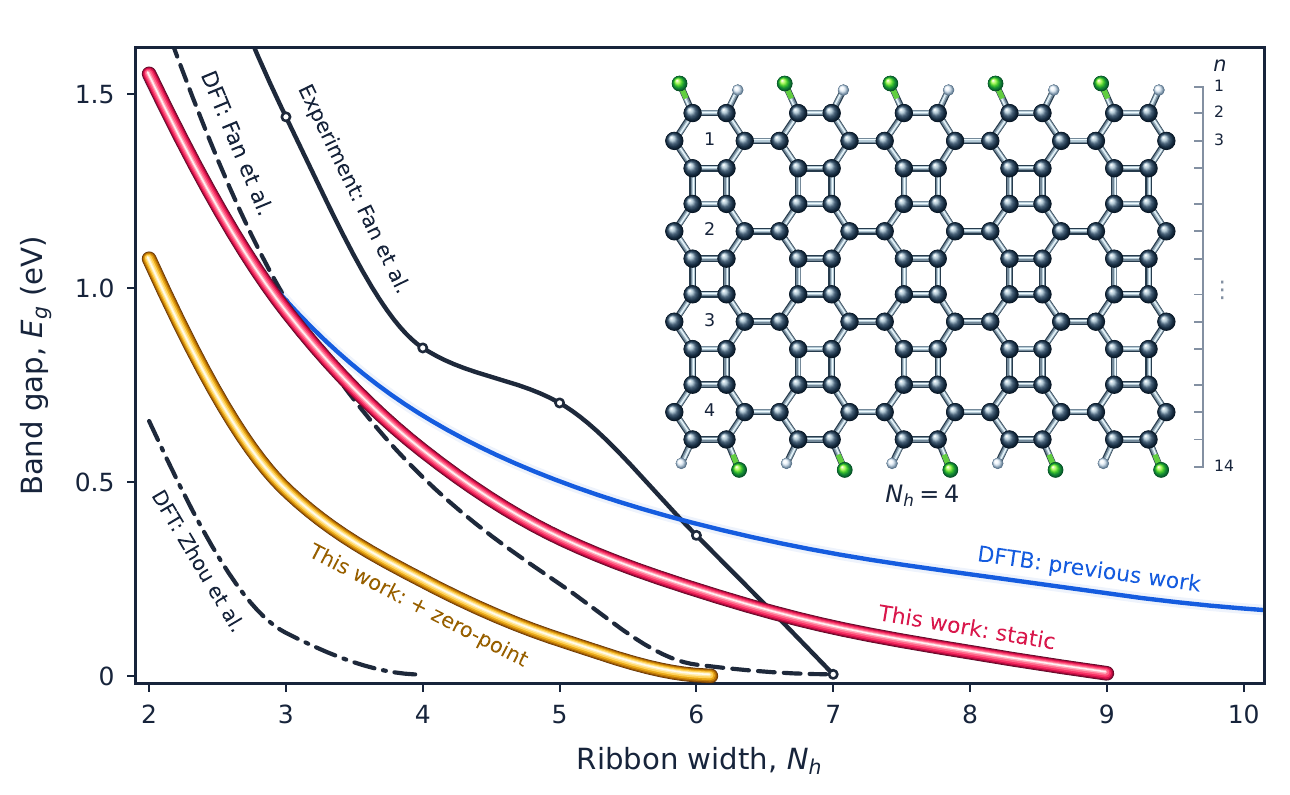}
\caption{Width-dependent band gap. Red and gold curves show the present static-lattice and zero-point-renormalized results, respectively. Black solid and dashed curves are experiment and DFT from Ref.~\onlinecite{Fan2021}; black dash-dotted and blue curves are DFT from Ref.~\onlinecite{Zhou2024} and earlier DFTB results~\cite{Katkov2026}. The inset defines $N_h$ for a ribbon with $N_h=4$.}
\label{fig:1}
\end{figure}

\textit{Computational framework.---}Ribbon width is specified by the number $N_h$ of hexagons across the ribbon (see Fig.~\ref{fig:1}). Periodic boundary conditions are imposed along its axis. Structures and phonons are treated using the PET-MAD-S machine-learned potential from UPET~\cite{Deringer2019,Bigi2026} and Phonopy~\cite{phonopy2015,phonopy2023,phonopy20232}. For supported ribbons, a nearly commensurate supercell contains approximately $2\times10^3$ atoms with longitudinal mismatch below $0.3\%$; hBN atoms are fixed during structural relaxation. Electronic energies are calculated using density-functional tight binding (DFTB)~\cite{Hourahine2020} with PTBP parameters~\cite{Cui2024PTBP}, without self-consistent charge iterations. The electronic calculation retains the ribbon atoms only; the substrate enters through the structural and vibrational model. 

PET-MAD-S was selected by comparing equilibrium lattice parameters of the biphenylene network against a B3LYP reference calculation performed with the Vienna \textit{Ab initio} Simulation Package (VASP)~\cite{Kresse1996VASP} (further details are given in the Supplemental Material).

Frozen-phonon calculations~\cite{Capaz2005} use opposite displacements along each normal mode, with a maximum bond-length change of $0.0125$~\AA. Their symmetric combination isolates the even electronic response. Within the harmonic description,
\begin{align}
E_g(T)&=E_g^{\mathrm{stat}}+\sum_i W_i\left[n_i(T)+\tfrac12\right],\label{eq:gap}\\
\Delta E_g(T)&\equiv E_g(T)-E_g(0)=\sum_i W_i n_i(T),\label{eq:thermal}
\end{align}
where $n_i(T)=[\exp(\hbar\omega_i/k_{\mathrm B}T)-1]^{-1}$ and $W_i$ is the signed gap-renormalization weight of mode $i$. The zero-point correction is $\Delta E_g^{\mathrm{ZP}}=\frac12\sum_iW_i$.
Additional computational details are given in the Supplemental Material.

The zero-point correction includes all modes, whereas the thermal shift additionally weights them by the Bose occupation and is therefore dominated by low-frequency modes at low temperatures.
An additional temperature-dependent shift accounts for the same modes with weights corresponding to their Bose-statistical occupancy, which renders the contribution of low-frequency modes more significant at low temperatures. Thus, the mode dominating the initial temperature slope does not necessarily play the decisive role in the total zero-point fluctuation correction.

\textit{Electronic consequences of lattice motion.---}Figure~\ref{fig:1} first establishes the importance of zero-point motion. The static DFTB gap decreases approximately as $N_h^{-1}$ and closes near $N_h\simeq9$. Zero-point renormalization shifts this closure to $N_h\simeq6$, closer to the experimental crossover near $N_h\simeq7$~\cite{Fan2021}. The calculations compared in the figure differ quantitatively, so this shift should be understood within the present electronic-structure approximation. Nevertheless, it shows that quantum lattice motion can materially alter the width at which a ribbon becomes gapless.

\begin{figure}[!htbp]
\centering
\includegraphics[width=\columnwidth]{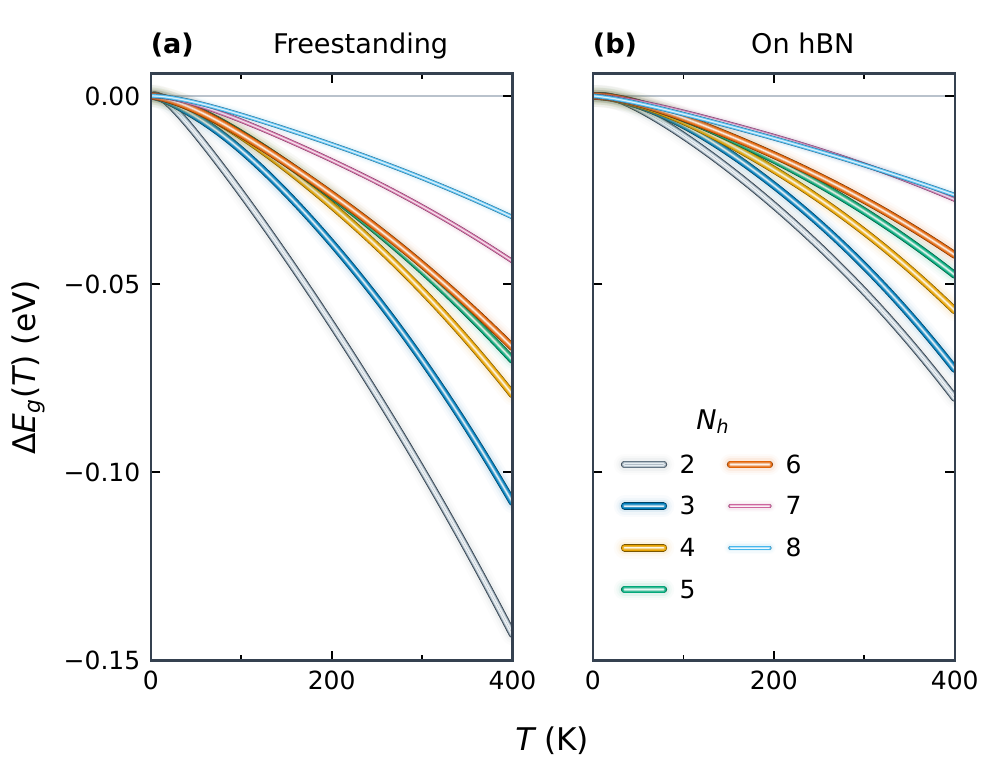}
\caption{Thermal band-edge shift $\Delta E_g(T)=E_g(T)-E_g(0)$ for (a) freestanding and (b) hBN-supported ribbons. Colors identify $N_h$. The reference $E_g(0)$ includes zero-point renormalization. For $N_h=7,8$, the curves track the change in band-edge separation after gap closure, rather than a finite insulating gap.}
\label{fig:2}
\end{figure}

Thermal occupation produces a further negative shift (see Fig.~\ref{fig:2}), whose magnitude decreases with width and remains smaller than the zero-point correction over the investigated temperature range. For $N_h=7,8$, the renormalized gap has already closed; the plotted quantity is the shift of the tracked band-edge separation. Although the substrate causes only a small change in the static gap ($\leq 0.01$~eV), it noticeably reduces the thermal response.

\begin{figure}[!htbp]
\centering
\includegraphics[width=\columnwidth]{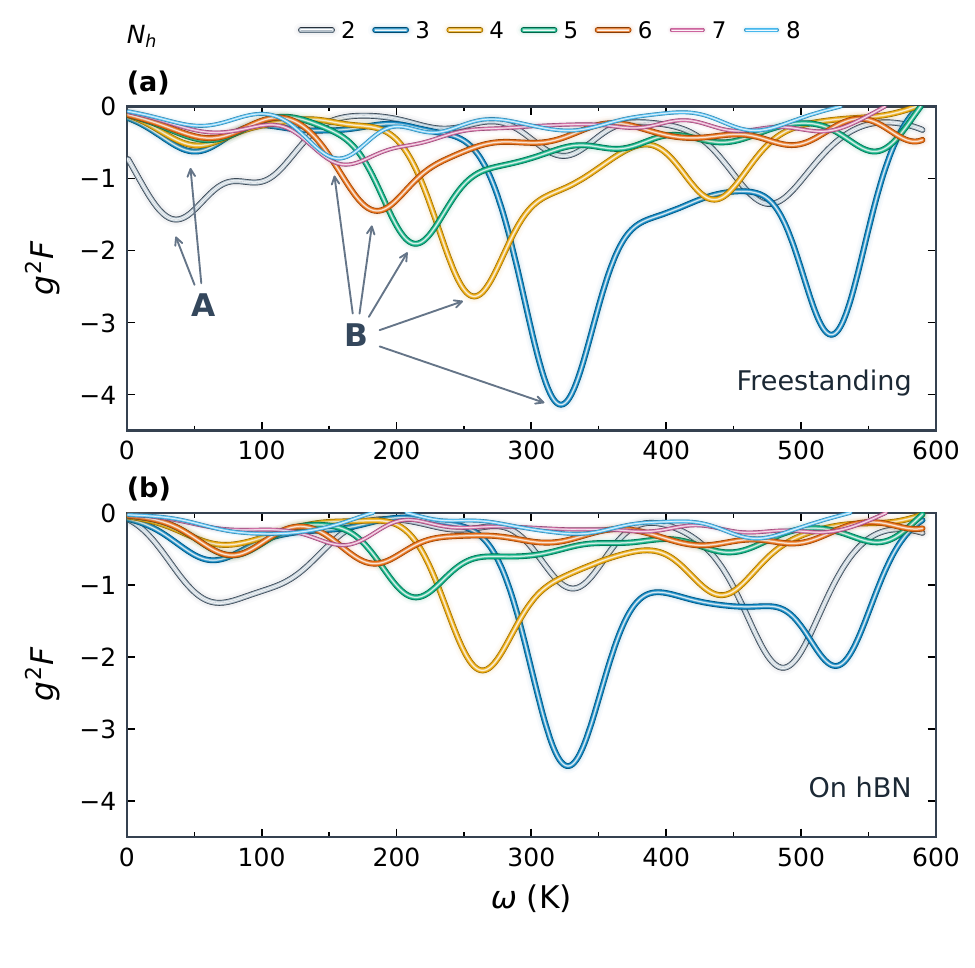}
\caption{Signed gap-response spectrum $g^2F(\omega)$ [Eq.~\eqref{eq:spectral}] for (a) freestanding and (b) hBN-supported ribbons. A and B mark the low-frequency and width-dependent groups of features. Colors label $N_h$. Frequencies are shown as $\hbar\omega/k_{\mathrm B}$, with Gaussian broadening $\sigma=23$ K.}
\label{fig:3}
\end{figure}

Because hBN electronic states are excluded from the calculation, the substrate-induced changes shown in Fig.~\ref{fig:2} arise from its effect on the ribbon geometry and vibrations.
To resolve this response by frequency, we use the signed gap-response spectrum
\begin{equation}
g^2F(\omega)=\sum_i W_i\delta(\omega-\omega_i).\label{eq:spectral}
\end{equation}
In Fig.~\ref{fig:3}, each $\delta$ function is replaced by a normalized Gaussian with $\sigma=23$~K.
Before broadening, Eq.~\eqref{eq:spectral} makes the connection to the gap shift explicit:
\begin{equation}
\Delta E_g(T)=\int_0^\infty g^2F(\omega)n_{\mathrm B}(\omega,T)\,d\omega.
\label{eq:spectral_integral}
\end{equation}

As seen in Fig.~\ref{fig:3},
two types of spectral features, labeled A and B, provide the main contributions to the thermal gap shift. The low-frequency feature A lies at $40$--$60$~K for freestanding ribbons and approximately $50$--$70$~K for ribbons on hBN. Its position depends only weakly on $N_h$, whereas its amplitude decreases with increasing width. Along the sequence $N_h=3$~to~$8$, feature B both weakens and shifts to lower frequencies. Both groups persist on the substrate, but their amplitudes generally decrease, consistent with the weaker thermal narrowing of the gap.

The low frequencies of the modes in feature A lead to their thermal occupation already at relatively low temperatures. This feature determines the initial temperature response and retains a substantial contribution up to room temperature. As the temperature rises, the role of the higher-frequency modes in feature B increases, and near room temperature their contribution becomes comparable to that of A.

\begin{figure}[!htbp]
\centering
\includegraphics[width=\columnwidth]{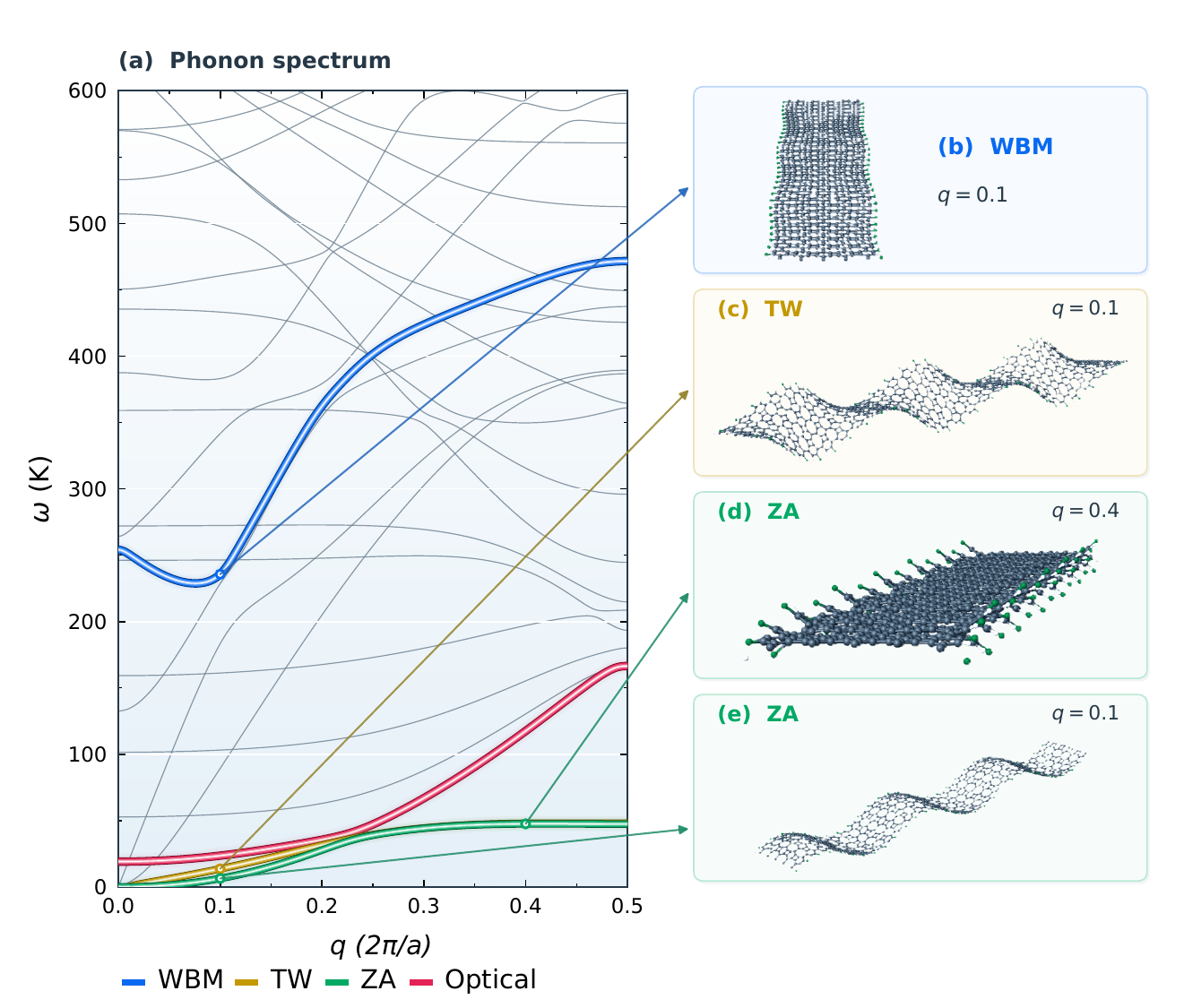}
\caption{(a) Phonon spectrum of a freestanding ribbon with $N_h=5$, calculated within the tight-binding approximation. The colored branches determine features A and B of the spectral function $g^2F(\omega)$. (b)--(e) Displacement patterns of selected modes at the indicated values of $q$: WBM is the width-breathing mode, TW is the twisting mode, and ZA is the flexural acoustic mode. The displacement amplitudes are magnified for clarity.}
\label{fig:4}
\end{figure}

\textit{Breathing modes and nearly flat branches.---}
The relation between the spectral features and specific types of motion is established by analyzing the weights $W_i$ together with the phonon eigenvectors. Figure~\ref{fig:4} illustrates this analysis using the spectrum of a freestanding ribbon with $N_h=5$.

Feature B is formed predominantly by the U-shaped segment of the width-breathing branch (WBM, blue curve). Its electronic response is small directly at the $\Gamma$ point and increases markedly at small but finite $q$. To interpret this difference, consider the symmetric combination of the responses to opposite displacements, 
\begin{equation}
    W_i=\left[\Delta E_g(+c_iQ_i)+\Delta E_g(-c_iQ_i)\right]/(2c_i^2),
\end{equation}
where $\Delta E_g(\pm c_iQ_i)$ denotes the change in the gap under a frozen displacement relative to the equilibrium configuration. The factor $c_i$ specifies the trial amplitude for the adopted normalization of $Q_i$~\cite{Cohen1986}.

At the $\Gamma$ point, the opposite displacements correspond to expansion and contraction of the ribbon. Symmetrization removes the odd, in particular linear, part of the response; the resulting small $W_i$ indicates a weak residual even component for this mode. At finite $q$, a spatially nonuniform transverse distortion appears. For the modes considered here, reversing the sign of the displacement is nearly equivalent to shifting the deformation profile along the ribbon. The two frozen distortions therefore produce similar gap changes of the same sign and a substantial even response.

Increasing the ribbon width softens the transverse breathing motion, which explains the shift of feature B to lower frequencies. The decrease in its amplitude is accompanied by a reduction in the corresponding mode weights and is consistent with a weaker relative transverse deformation for comparable atomic displacements.

Feature A consists of comparable contributions from three branches: a flexural optical branch (red), the twisting branch TW (gold), and the flexural acoustic branch ZA (green). Their weakly dispersive segments provide a high density of states within a narrow frequency interval, while the appreciable weights $W_i$ make these states effective contributors to the gap renormalization. Thus, the high phonon density of states acts together with the sensitivity of the electronic spectrum to the corresponding distortions.

The behavior of the TW and ZA branches is particularly informative. At small $q$, they describe ordinary extended vibrations of the ribbon. Above a characteristic wave vector $q_c$, both branches become nearly flat; for $N_h>3$, their frequencies nearly coincide. In the narrowest ribbons, $N_h=2,3$, weakly dispersive segments also form but remain split by approximately $10$~K. It is interesting to identify the type of motion in these nearly dispersionless states.

\textit{The edge as a system of atomic resonators.---}
Analysis of the eigenvectors shows that the flattening of TW and ZA is accompanied by a transfer of most of the displacement amplitude to the H--F edge complexes. The H and F atoms move in phase, whereas the displacements of the central part of the carbon backbone are strongly suppressed. The ZA branch evolves into oscillations of the opposite edges in phase, while TW evolves into oscillations with a phase difference of $\pi$. The disappearance of an appreciable splitting between these branches for $N_h>3$ is consistent with weaker dynamical coupling between the edges.

\begin{figure}[!htbp]
\centering
\includegraphics[width=\columnwidth]{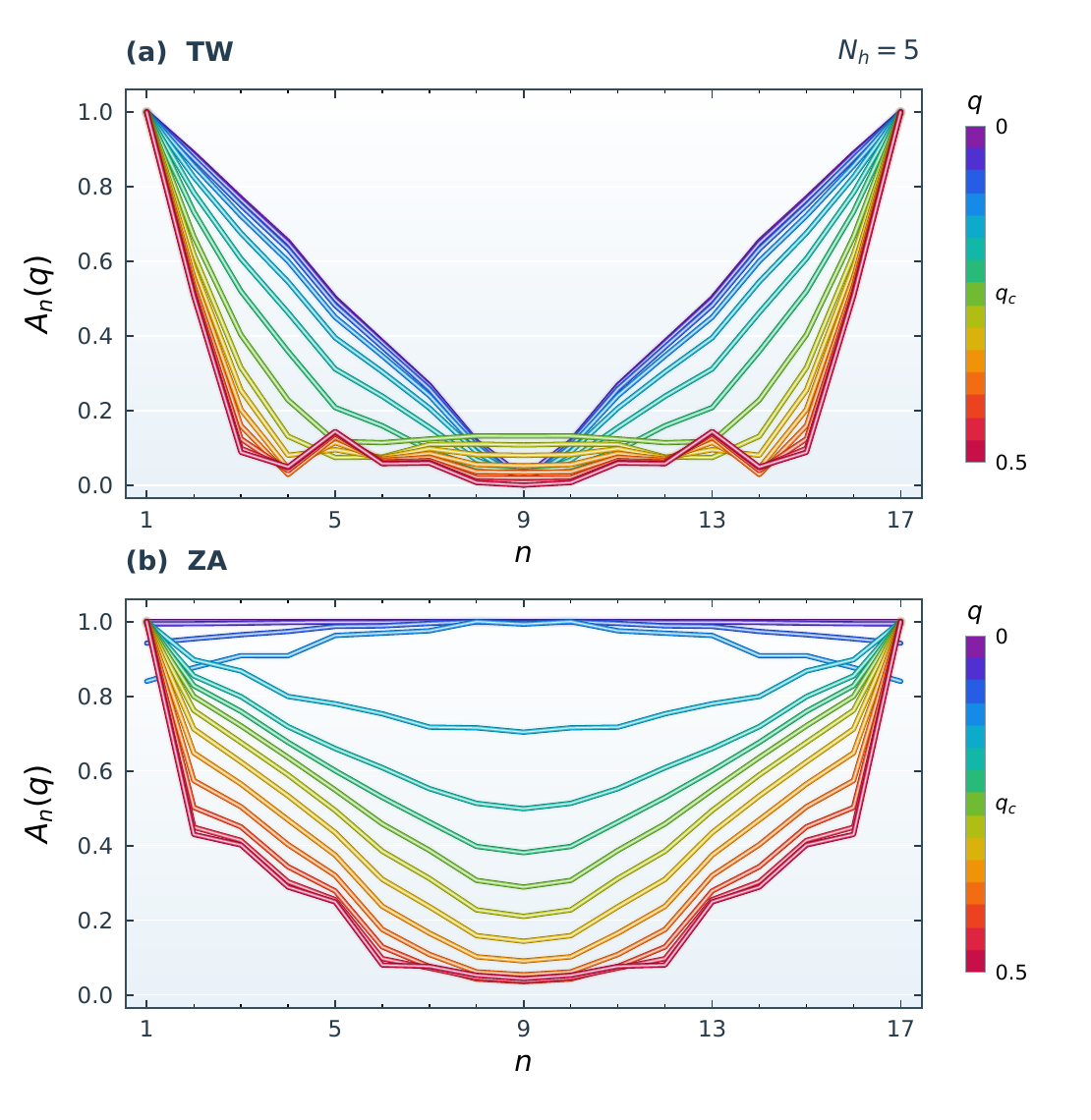}
\caption{Normalized transverse profiles of the atomic displacement amplitudes $A_n(q)$ for the TW (a) and ZA (b) branches of a freestanding ribbon with $N_h=5$. The index $n$ labels the atomic rows across the ribbon, and color denotes the wave vector $q$. The value $q_c$ corresponds to the crossover from motion distributed across the ribbon width to predominantly edge-localized motion.}
\label{fig:5}
\end{figure}

The rearrangement of the transverse profile is shown in Fig.~\ref{fig:5}. For atom $i$, the displacement amplitude is defined using the polarization vector:
$
u_i(q)=|\mathbf e_i(q)|/\sqrt{m_i}.
$
For each transverse row (see Fig.~\ref{fig:1})  the amplitudes of the two atoms belonging to the same unit cell are summed,
\begin{equation}
S_n(q)=u_{n,1}(q)+u_{n,2}(q),~
A_n(q)=\frac{S_n(q)}{\max_m S_m(q)}.
\end{equation}
In particular, for the edge row, $S_1=u_{1,\mathrm H}+u_{1,\mathrm F}$. The normalization is performed separately at each $q$.

As $q$ increases, the profile changes continuously from motion distributed across the ribbon width to predominantly edge-localized motion.  
The local character of the resonance explains the weak dependence of the frequency of the flat TW and ZA segments on ribbon width and, therefore, the stable position of feature A. The decrease in its spectral weight is consistent with the weaker relative influence of the edge region on the electronic response as $N_h$ increases. In control calculations for ribbons with H--H edge complexes, the characteristic combination of pronounced branch flattening and edge localization is absent, and the temperature dependence of $E_g$ is weaker~\cite{SupplementalMaterial}.

Numerically varying the F mass provides an independent diagnostic. Near its physical value, the flat-branch frequency follows
\begin{equation}
\omega_{\mathrm{flat}}\simeq\sqrt{K/(m_0+m_x)},\label{eq:resonance}
\end{equation}
where $m_x$ is the varied mass, $m_0$ is a few atomic mass units, and $K\sim1$--$3$ N/m. These values are characteristic of a soft transverse coordinate and are orders of magnitude smaller than typical bond-stretching force constants. Together with the eigenvectors, this scaling supports a local bending resonance coupled to the backbone.

Mass tuning is useful because it changes the inertia of the proposed resonator without changing the force constants. Agreement with Eq.~\eqref{eq:resonance} over the examined neighborhood therefore tests the dynamical assignment separately from a change in chemical bonding. The offset $m_0$ represents an effective contribution of the surrounding moving atoms, rather than an additional species at the edge. Replacing F chemically is a different operation: both mass and bonding may change.
 
A \textit{minimal mechanical model} reproduces this crossover and shows that it requires a low-frequency edge resonance overlapping a ribbon branch and sufficiently weak longitudinal coupling between edge complexes~\cite{SupplementalMaterial}.
Stronger longitudinal coupling gives the edge subsystem its own dispersion and can shift it away from resonance. The larger effective mass of H--F, combined with a soft bending coordinate, makes low-frequency matching possible. This provides an atomic counterpart of a tuned mass damper~\cite{Warburton1982}: a resonator carries a large amplitude while motion of the supporting structure is reduced. Here the analogy concerns harmonic mode shapes and does not invoke dissipative damping. Unlike the mere existence of an edge phonon, as in Tamm states~\cite{Savin2010}, the wave-vector-dependent conversion of a backbone mode into an edge resonance identifies the relevant dynamical mechanism.

For this mechanism to be realized, three conditions must be met: a low edge-resonance frequency, spectral overlap with an extended phonon branch of the ribbon, and weak dispersion of the edge-resonance branch.
These conditions can be varied by changing the end groups and the nature of their linkage to the main chain.
It suggests tuning terminal groups and local coupling parameters while monitoring the collective spectrum of the ribbon after each modification.
Nearly flat branches may also affect thermal transport, as demonstrated for resonant nanophononic architectures~\cite{Davis2014,Xiong2016}. Establishing that consequence here requires lifetimes and a transport calculation beyond the present harmonic gap analysis. The demonstrated electronic response already provides a separate reason to engineer these modes, even without assuming a reduction of thermal conductivity.
 
Variations in the width and structure of the edges provide complementary ways to investigate this mechanism. Increasing width changes transverse confinement and the separation between the edges, whereas changing the terminal complex modifies the local resonant coordinate.
The specific dependence of A and B on the width makes it possible to distinguish local edge oscillations from collective transverse modes.
A similar mechanism may occur in other nanoribbons whose chemically modified edges support low-frequency local resonances. Collectively, these criteria could underpin a specific strategy for the chemical modification of nanoribbon edges.

\textit{Conclusion.---}Biphenylene nanoribbon edges can act as an intrinsic atomic resonator array. Their local bending dynamics flatten low-frequency phonon branches, suppress backbone motion, and contribute to thermal band-gap narrowing. Zero-point motion shifts gap closure toward the observed width range, while hBN support preserves the principal spectral features with a reduced thermal response. Edge mass, bonding stiffness, and coupling along the edge therefore offer physical control parameters connecting local chemistry to collective vibrations and electronic structure. The key conclusion is that the atomic edge can serve as an active element in shaping the phonon band structure: a small set of edge bonds can alter the response of the entire nanoribbon.

\nocite{Kresse1999PAW,Wood2025,Sadasivam2014}
\bibliographystyle{apsrev4-2}
\bibliography{references1}
\end{document}

\title{Supplemental Material for\\
Atomic Edge Resonators Flatten Phonon Bands in Biphenylene Nanoribbons}
\author{V.~L.~Katkov}
\affiliation{Bogoliubov Laboratory of Theoretical Physics, Joint Institute for Nuclear Research, Dubna, Russia}
\author{N.~M.~Chtchelkatchev}
\affiliation{Bogoliubov Laboratory of Theoretical Physics, Joint Institute for Nuclear Research, Dubna, Russia}
\author{V.~A.~Osipov}
\affiliation{Bogoliubov Laboratory of Theoretical Physics, Joint Institute for Nuclear Research, Dubna, Russia}
\maketitle

This Supplemental Material describes the computational framework and the mode-resolved electronic response used in the Letter and presents a minimal mechanical model of resonance-induced edge localization.

\section{Minimal Effective Model for Resonance-Induced Edge Phonon Localization}

\subsection{Description of the system}

The present model describes the dynamics of a one-dimensional discrete mechanical system consisting of a semi-infinite atomic chain with a concentrated edge mass and a local elastic foundation.

\begin{figure}[htbp]
    \centering
    \includegraphics[width=0.62\textwidth]{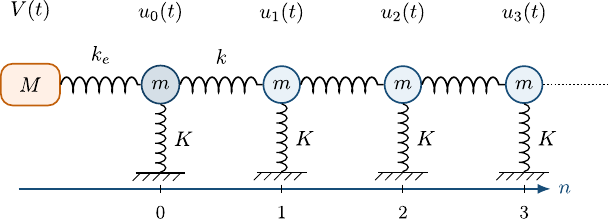}
    \caption{A semi-infinite atomic chain with an edge mass $M$. A spring of stiffness $k_e$ connects the edge mass to the first backbone atom, springs of stiffness $k$ connect neighboring backbone atoms, and $K$ specifies the local restoring stiffness.}
    \label{fig:single_chain}
\end{figure}

The system comprises three key components: an edge element, a coupling element, and a semi-infinite backbone. The edge element is represented by a concentrated mass $M$, whose dynamics is described by the coordinate $V(t)$. The coupling element is an elastic spring of stiffness $k_e$ connecting the edge mass to the first atom of the backbone. The semi-infinite backbone itself is an ordered array of atoms whose coordinates are denoted by $u_n(t)$ for $n\geq0$ (see Fig.~\ref{fig:single_chain}).

\subsection{Model parameters and variables}

The following dynamical variables are used to describe the system:
\begin{itemize}
    \item $V(t)$ --- the generalized coordinate (displacement) of the edge mass $M$ at time $t$;
    \item $u_0(t)$ --- the generalized coordinate (displacement) of the zeroth (edge) backbone atom;
    \item $u_n(t)$ --- the generalized coordinate (displacement) of the $n$th backbone atom for $n\geq1$.
\end{itemize}

The physical properties of the elements and their interactions are specified by the following parameters:
\begin{itemize}
    \item $M$ --- mass of the edge element, kg;
    \item $m$ --- mass of each individual backbone atom, kg;
    \item $k_e$ --- stiffness of the coupling spring between the mass $M$ and the zeroth atom $u_0$, N/m;
    \item $k$ --- longitudinal stiffness of the springs connecting neighboring backbone atoms, N/m;
    \item $K$ --- effective transverse (local) stiffness of the foundation (substrate), which keeps each backbone atom near its equilibrium position, N/m.
\end{itemize}

\subsection{Mathematical formulation}

The dynamics of the system is described by a closed set of second-order ordinary differential equations.

The motion of the edge mass is governed by
\begin{equation}
M \ddot{V}(t) = -k_e \Big(V(t) - u_0(t)\Big).
\end{equation}
The dynamics of the zeroth (boundary) backbone atom includes the combined effects of the substrate, the edge spring, and the first interatomic bond of the chain:
\begin{equation}
m \ddot{u}_0(t) = -K u_0(t) - k \Big(u_0(t) - u_1(t)\Big) - k_e \Big(u_0(t) - V(t)\Big).
\end{equation}
For all subsequent interior backbone atoms ($n\geq1$), the equation of motion has the standard form for a discrete chain on an elastic foundation:
\begin{equation}
m \ddot{u}_n(t) = -K u_n(t) - k \Big(2u_n(t) - u_{n-1}(t) - u_{n+1}(t)\Big), \quad n \ge 1.
\end{equation}

\subsection{Dynamical features and boundary conditions}

The boundary atom $u_0$ acts as a coupling node and is affected by the edge mass $V$, the first chain atom $u_1$, and the external foundation with stiffness $K$.

The presence of an effective transverse stiffness $K>0$ opens a spectral (dispersion) gap. Consequently, propagating waves in the backbone have a lower frequency threshold: disturbances with frequencies below the foundation cutoff frequency $\omega_{\min}=\sqrt{K/m}$ decay and cannot propagate along the backbone.

\subsection{Frequency-domain representation and dynamical matrix}

To analyze steady-state wave motion and the spectral properties of the backbone, we pass from the time domain to the frequency domain. The solution is sought as normal modes with harmonic time dependence,
\begin{equation}
V(t)=V e^{-i\omega t},
\qquad
u_n(t)=U_n e^{-i\omega t},
\qquad n\geq 0,
\end{equation}
where $V$ and $U_n$ are complex amplitudes and $\omega$ is the angular frequency.

Substitution of this harmonic ansatz and $\ddot{u}_n(t)=-\omega^2U_ne^{-i\omega t}$ into the equations of motion for the interior backbone atoms ($n\geq1$) yields an algebraic system for the amplitudes $U_n$:
\begin{equation}\label{eq:5}
\Big(K + 2k - m\omega^2\Big) U_n - k U_{n-1} - k U_{n+1} = 0, \quad n \ge 1.
\end{equation}

For the homogeneous semi-infinite tail $U_1,U_2,\dots$, introduce the vector $\mathbf{U}_{\infty}=(U_1,U_2,U_3,\dots)^T$. Its dynamics is governed by $\mathbf{D}_{\infty}(\omega)\mathbf{U}_{\infty}=\mathbf{0}$, with the tridiagonal matrix

\begin{equation}
\mathbf{D}_{\infty}(\omega) = 
\begin{pmatrix}
a(\omega) & -k & 0 & 0 & \dots \\
-k & a(\omega) & -k & 0 & \dots \\
0 & -k & a(\omega) & -k & \dots \\
0 & 0 & -k & a(\omega) & \dots \\
\vdots & \vdots & \vdots & \vdots & \ddots
\end{pmatrix}.
\end{equation}

The components of this matrix are defined as follows:

\begin{itemize}
    \item The \textbf{diagonal elements} $a(\omega)$ describe the local dynamical response of an individual node, including its inertia, its coupling to the foundation, and the two adjacent interatomic springs:
    \begin{equation}
    a(\omega) = K + 2k - m\omega^2.
    \end{equation}
    \item The \textbf{off-diagonal elements} are equal to $-k$ and represent the pairwise interaction of the $n$th atom with its nearest neighbors, $(n-1)$ and $(n+1)$.
\end{itemize}
To include external excitation, introduce the vector of complex amplitudes of the external forces,
\begin{equation}
\mathbf{F} =
\begin{pmatrix}
F_1 & F_2 & F_3 & \dots
\end{pmatrix}^{T},
\end{equation}
where $F_n$ is the complex amplitude of a harmonic external force applied to the $n$th atom of the backbone. The equation for forced oscillations in the frequency domain then takes the matrix form
\begin{equation}
\mathbf{D}_{\infty}(\omega)\mathbf{U}=\mathbf{F},
\end{equation}
or, equivalently,
\begin{equation}
\mathbf{U}=\mathbf{D}_{\infty}^{-1}(\omega)\mathbf{F},
\end{equation}
provided that the inverse operator $\mathbf{D}_{\infty}^{-1}(\omega)$ exists. In the absence of external excitation, $\mathbf{F}=\mathbf{0}$, this equation becomes the homogeneous spectral problem
\begin{equation}
    \mathbf{D}_{\infty}(\omega)\mathbf{U}=\mathbf{0}.
\end{equation}

\subsection{Matrix and surface Green's functions}

We next use the Green's-function technique~\cite{Sadasivam2014}. For the homogeneous semi-infinite tail, define the \textbf{matrix Green's function} $\mathbf{G}(\omega)=\mathbf{D}_{\infty}^{-1}(\omega)$, which satisfies
\begin{equation}
\mathbf{D}_{\infty}(\omega)\mathbf{G}(\omega)=\mathbf{I},
\end{equation}
where $\mathbf{I}$ is the infinite identity matrix.

Physically, the element $G_{n,m}(\omega)$ is the complex displacement amplitude of the $n$th atom produced by a unit-amplitude harmonic force of frequency $\omega$ applied exclusively to the $m$th atom.

If an arbitrary vector of external harmonic forces $\mathbf{F}(\omega)=(F_1,F_2,F_3,\dots)^T$ acts on the backbone, the full displacement vector is obtained directly from the Green's function:
\begin{equation}
\mathbf{U} = \mathbf{G}(\omega) \mathbf{F}(\omega).
\end{equation}

The surface Green's function is defined as the first diagonal element of the full Green's-function matrix:
\begin{equation}
g(\omega) \equiv G_{1,1}(\omega) = \big[\mathbf{D}^{-1}_{\infty}(\omega)\big]_{1,1}.
\end{equation}

The surface Green's function $g(\omega)$ gives the response of the backbone boundary ($n=1$) to a localized unit force $F_1$, and therefore relates the displacement amplitude of the first atom directly to the applied force: $U_1=g(\omega)F_1$. Knowledge of $g(\omega)$ makes it possible to eliminate all internal variables $U_2,U_3,\dots$ from the equations of motion and reduce the infinite backbone system to a single effective equation for its boundary atom.

\subsection{Derivation of the recursive equation using block matrices}

The recursive property of the surface Green's function follows directly from the block structure of the infinite dynamical matrix $\mathbf{D}_{\infty}(\omega)$. Isolate the first node of the homogeneous tail and write the operator in block-matrix form:
\begin{equation}
\mathbf{D}_{\infty}(\omega) = 
\begin{pmatrix}
a(\omega) & \mathbf{B}^T \\
\mathbf{B} & \mathbf{D}_{\mathrm{\infty}}(\omega)
\end{pmatrix},
\end{equation}
where $\mathbf{B}=(-k,0,0,\dots)^T$ is the column vector that couples the first node to the remaining backbone, while $\mathbf{D}_{\mathrm{\infty}}(\omega)$ is the submatrix of the remaining semi-infinite tail. Because the semi-infinite chain is homogeneous and self-similar, this submatrix is structurally identical to the matrix on the left-hand side.

The full Green's-function matrix $\mathbf{G}(\omega)=\mathbf{D}_{\infty}^{-1}(\omega)$ can likewise be written in block form:
\begin{equation}
\mathbf{G}(\omega) = 
\begin{pmatrix}
G_{1,1}(\omega) & \mathbf{G}_{1,\infty}(\omega) \\
\mathbf{G}_{\infty,1}(\omega) & \mathbf{G}_{\infty,\infty}(\omega)
\end{pmatrix}.
\end{equation}

The block-inversion formula, expressed through the Schur complement, is
\begin{equation}
\begin{pmatrix} A & B \\ C & D \end{pmatrix}^{-1} = \begin{pmatrix} (A - B D^{-1} C)^{-1} & \dots \\ \dots & \dots \end{pmatrix}.
\end{equation}
Matching the blocks and selecting the upper-left element, which here is the scalar $G_{1,1}(\omega)\equiv g(\omega)$, gives the exact expression
\begin{equation}
g(\omega) = \Big( a(\omega) - \mathbf{B}^T \mathbf{D}_{\mathrm{\infty}}^{-1}(\omega) \mathbf{B} \Big)^{-1}.
\end{equation}

Direct multiplication, using the structure of $\mathbf{B}$, reduces the quadratic form in parentheses to the first diagonal element of the inverse tail submatrix, $[\mathbf{D}_{\mathrm{\infty}}^{-1}(\omega)]_{1,1}$:
\begin{equation}
\mathbf{B}^T \mathbf{D}_{\mathrm{\infty}}^{-1}(\omega) \mathbf{B} = 
\begin{pmatrix} -k & 0 & 0 & \dots \end{pmatrix}
\mathbf{D}_{\mathrm{\infty}}^{-1}(\omega)
\begin{pmatrix} -k \\ 0 \\ 0 \\ \vdots \end{pmatrix} = 
k^2 \big[\mathbf{D}_{\mathrm{\infty}}^{-1}(\omega)\big]_{1,1}.
\end{equation}

Owing to the self-similarity of the semi-infinite chain, the local response at the boundary of the isolated tail is identical to the response at the boundary of the full system: $[\mathbf{D}_{\mathrm{\infty}}^{-1}(\omega)]_{1,1}=g(\omega)$. This closes the matrix relation and gives the algebraic equation
\begin{equation}
g(\omega) = \frac{1}{a(\omega) - k^2 g(\omega)}.
\end{equation}

This equation is equivalent to the quadratic equation $k^2g^2(\omega)-a(\omega)g(\omega)+1=0$, whose two analytic branches are
\begin{equation}
g_{1,2}(\omega) = \frac{a(\omega) \pm \sqrt{a^2(\omega) - 4k^2}}{2k^2}.
\end{equation}
The physically admissible branch is selected by the boundary conditions of the semi-infinite chain and by the corresponding spatial behavior of the solution. 

\subsection{Spatial decay factor and its relation to the surface Green's function}

To analyze localized (nonpropagating) backbone states, consider the spatial variation of the oscillation amplitude with distance from the boundary into the chain. We seek the complex amplitudes as a geometric progression,
\begin{equation}
U_n = U_1 \lambda^{n-1}, \quad n = 1,2,3,\dots,
\end{equation}
where $\lambda$ is the dimensionless localization factor, defined as the ratio of the displacement amplitudes of neighboring atoms:
\begin{equation}
\lambda = \frac{U_{n+1}}{U_n}.
\end{equation}
In particular,
\begin{equation}
\lambda = \frac{U_2}{U_1}.
\end{equation}
The amplitude decays away from the boundary when
\begin{equation}
|\lambda| < 1.
\end{equation}
For $0<\lambda<1$, the oscillation amplitude decreases into the chain without a sign change between neighboring atomic displacements. For $-1<\lambda<0$, its magnitude also decays exponentially, but the sign of the displacement alternates from atom to atom.

The characteristic localization length $\xi$, expressed in lattice-spacing units, is
\begin{equation}
\xi = -\frac{1}{\ln |\lambda|}.
\end{equation}

We now establish the relation between the localization factor $\lambda$ and the surface Green's function $g(\omega)$. Apply an external harmonic force with complex amplitude $F$ to the first atom of the semi-infinite backbone. In the frequency domain, the equation for this atom is
\begin{equation}
a(\omega) U_1 - k U_2 = F.
\end{equation}
By definition of the surface Green's function,
\begin{equation}
U_1 = g(\omega) F,
\end{equation}
and hence
\begin{equation}
F = \frac{U_1}{g(\omega)}.
\end{equation}
Substitution into the equation for the first atom gives
\begin{equation}
a(\omega) U_1 - k U_2 = \frac{U_1}{g(\omega)}.
\end{equation}
After division by $U_1$,
\begin{equation}
a(\omega) - k\frac{U_2}{U_1} = \frac{1}{g(\omega)}.
\end{equation}
Since
\begin{equation}
\frac{U_2}{U_1} = \lambda,
\end{equation}
we obtain
\begin{equation}
\frac{1}{g(\omega)} = a(\omega) - k\lambda.
\end{equation}
On the other hand, the Schur-complement construction gave the recurrence relation
\begin{equation}\label{eq:34}
\frac{1}{g(\omega)} = a(\omega) - k^2 g(\omega).
\end{equation}
Comparison of the two expressions gives the direct relation
\begin{equation}\label{eq:35}
\lambda = k g(\omega).
\end{equation}
The corresponding localization factors are therefore
\begin{equation}
\lambda_{\pm}(\omega)
=
\frac{a(\omega) \pm \sqrt{a^2(\omega)-4k^2}}{2k}.
\end{equation}
Their product equals unity:
\begin{equation}
\lambda_{+}(\omega)\lambda_{-}(\omega) = 1.
\end{equation}

For a localized state, the physically admissible root is the one for which the amplitude decays away from the boundary of the semi-infinite chain:
\begin{equation}
|\lambda(\omega)| < 1.
\end{equation}
The other root then satisfies $|\lambda(\omega)|>1$ and corresponds to a solution that grows as $n\to\infty$; it must therefore be discarded for a semi-infinite backbone. Using $\lambda=kg(\omega)$, the general condition selecting the physical branch of the surface Green's function is
\begin{equation}
|k g(\omega)|\leq 1.
\end{equation}
For $|kg|<1$, the solution decays into the chain. $|kg|=1$ corresponds to a nondecaying spatial profile.

\subsection{Reduction of the extended dynamical matrix to an effective two-component problem}

Consider the complete frequency-domain system consisting of the edge mass $V$, the boundary atom $U_0$, and the homogeneous semi-infinite backbone $U_1,U_2,\dots$. Introduce the full vector of complex amplitudes
\begin{equation}
\mathbf{U}_{\mathrm{full}}
=
\begin{pmatrix}
V & U_0 & U_1 & U_2 & \dots
\end{pmatrix}^{T}.
\end{equation}
In the absence of external excitation, the equations of motion of the complete system can be written in matrix form
\begin{equation}
\mathbf{D}_{\mathrm{full}}(\omega)
\mathbf{U}_{\mathrm{full}}
=
\mathbf{0},
\end{equation}
where the extended dynamical matrix is
\begin{equation}
\mathbf{D}_{\mathrm{full}}(\omega)
=
\begin{pmatrix}
k_e-M\omega^2
&
-k_e
&
0
&
0
&
\cdots
\\
-k_e
&
K+k+k_e-m\omega^2
&
-k
&
0
&
\cdots
\\
0
&
-k
&
a(\omega)
&
-k
&
\cdots
\\
0
&
0
&
-k
&
a(\omega)
&
\ddots
\\
\vdots
&
\vdots
&
\vdots
&
\ddots
&
\ddots
\end{pmatrix},
\end{equation}
where, as before,
\begin{equation}
a(\omega)=K+2k-m\omega^2.
\end{equation}

To eliminate the internal degrees of freedom, divide the full amplitude vector into boundary and internal parts:
\begin{equation}
\mathbf{U}_{b}
=
\begin{pmatrix}
V\\
U_0
\end{pmatrix},
\qquad
\mathbf{U}_{\infty}
=
\begin{pmatrix}
U_1\\
U_2\\
U_3\\
\vdots
\end{pmatrix}.
\end{equation}
The extended dynamical matrix can then be written in block form,
\begin{equation}
\mathbf{D}_{\mathrm{full}}(\omega)
=
\begin{pmatrix}
\mathbf{D}_{b}(\omega)
&
\mathbf{C}^{T}
\\
\mathbf{C}
&
\mathbf{D}_{\infty}(\omega)
\end{pmatrix},
\end{equation}
where
\begin{equation}
\mathbf{D}_{b}(\omega)
=
\begin{pmatrix}
k_e-M\omega^2
&
-k_e
\\
-k_e
&
K+k+k_e-m\omega^2
\end{pmatrix}
\end{equation}
describes the two boundary degrees of freedom. The coupling matrix is
\begin{equation}
\mathbf{C}
=
\begin{pmatrix}
0 & -k\\
0 & 0\\
0 & 0\\
\vdots & \vdots
\end{pmatrix}.
\end{equation}
The full system of equations becomes
\begin{equation}
\begin{pmatrix}
\mathbf{D}_{b}
&
\mathbf{C}^{T}
\\
\mathbf{C}
&
\mathbf{D}_{\infty}
\end{pmatrix}
\begin{pmatrix}
\mathbf{U}_{b}\\
\mathbf{U}_{\infty}
\end{pmatrix}
=
\begin{pmatrix}
\mathbf{0}\\
\mathbf{0}
\end{pmatrix}.
\end{equation}
The second block equation gives
\begin{equation}
\mathbf{U}_{\infty}
=
-\mathbf{D}_{\infty}^{-1}(\omega)
\mathbf{C}\mathbf{U}_{b}.
\end{equation}
Substituting this expression into the first block equation eliminates the infinite set of internal degrees of freedom:
\begin{equation}
\left[
\mathbf{D}_{b}(\omega)
-
\mathbf{C}^{T}
\mathbf{D}_{\infty}^{-1}(\omega)
\mathbf{C}
\right]
\mathbf{U}_{b}
=
\mathbf{0}.
\end{equation}
By definition, the surface Green's function of the semi-infinite tail is
\begin{equation}
g(\omega)
=
\left[
\mathbf{D}_{\infty}^{-1}(\omega)
\right]_{1,1}.
\end{equation}
Because the tail is coupled to the boundary only through its first atom, the matrix product in the Schur complement is
\begin{equation}
\mathbf{C}^{T}
\mathbf{D}_{\infty}^{-1}(\omega)
\mathbf{C}
=
\begin{pmatrix}
0 & 0\\
0 & k^2 g(\omega)
\end{pmatrix}.
\end{equation}
Thus, the original infinite-dimensional problem is reduced exactly to the effective two-component system
\begin{equation}
\mathbf{D}_{\mathrm{eff}}(\omega)
\begin{pmatrix}
V\\
U_0
\end{pmatrix}
=
\begin{pmatrix}
0\\
0
\end{pmatrix},
\end{equation}
where the effective dynamical matrix is
\begin{equation}
\mathbf{D}_{\mathrm{eff}}(\omega)
=
\begin{pmatrix}
k_e-M\omega^2
&
-k_e
\\
-k_e
&
K+k+k_e-m\omega^2-k^2g(\omega)
\end{pmatrix}.
\end{equation}
Introduce the effective dynamic stiffness of the boundary,
\begin{equation}
D_s(\omega)
=
K+k-m\omega^2-k^2g(\omega).
\end{equation}
Using the relation obtained above,
\begin{equation}
\lambda(\omega)=kg(\omega),
\end{equation}
it may also be written as
\begin{equation}
D_s(\omega)
=
K+k-m\omega^2-k\lambda(\omega).
\end{equation}
The effective dynamical matrix then takes the compact form
\begin{equation}
\mathbf{D}_{\mathrm{eff}}(\omega)
=
\begin{pmatrix}
k_e-M\omega^2
&
-k_e
\\
-k_e
&
k_e+D_s(\omega)
\end{pmatrix}.
\end{equation}
A nontrivial solution of the homogeneous system exists when
\begin{equation}
\det  [\mathbf{D}_{\mathrm{eff}}(\omega)]=0,
\end{equation}
which gives the following equation for $\omega$:
\begin{equation}
\big(k_e-M\omega^2\big)
\big[k_e+D_s(\omega)\big]
-k_e^2
=0.
\end{equation}
This equation determines the eigenfrequencies of the coupled motion of the edge mass and the boundary atom, including the exact dynamical effect of the entire semi-infinite backbone. Localized modes must additionally satisfy the spatial-decay condition
\begin{equation}
|\lambda(\omega)|<1.
\end{equation}
Because the boundary operator $D_s(\omega)$ depends nonlinearly on frequency through the surface Green's function $g(\omega)$, or equivalently through the decay factor $\lambda(\omega)$, this equation is strongly nonlinear.

\subsection{Antiresonance of the edge atom}

The first and second equations of the effective two-component system are
\begin{align}
\big(k_e-M\omega^2\big)V-k_eU_0&=0,\\
-k_eV+\big[k_e+D_s(\omega)\big]U_0&=0.
\end{align}
They give two equivalent expressions for the amplitude ratio:
\begin{equation}
\frac{U_0}{V}
=
\frac{k_e-M\omega^2}{k_e}
=
1-\frac{\omega^2}{\omega_0^2}
\end{equation}
and
\begin{equation}
\frac{U_0}{V}
=
\frac{k_e}{k_e+D_s(\omega)},
\end{equation}
where
\begin{equation}
\omega_0=\sqrt{\frac{k_e}{M}}
\end{equation}
is the resonance frequency of the attached edge resonator. Equating the two expressions gives
\begin{equation}
\omega^2
=
\omega_0^2
\frac{D_s(\omega)}{k_e+D_s(\omega)}.
\end{equation}
Thus, under the condition
\begin{equation}
|D_s(\omega)|\gg k_e,
\end{equation}
the eigenfrequency approaches the resonance frequency $\omega_0$. At the same time,
\begin{equation}
\frac{U_0}{V}
\simeq
\frac{k_e}{D_s(\omega)}
\longrightarrow0.
\end{equation}
Consequently, the amplitude of the boundary atom $U_0$ tends to zero while the amplitude of the attached mass remains finite. This regime has an antiresonant character: the resonant motion of the attached mass produces a compensating action that suppresses the oscillation of the node $U_0$.

For a localized solution, the amplitudes of the atoms in the semi-infinite backbone are related to the boundary amplitude by the following expression. Using Eq.~(\ref{eq:5}) for $n=1$ together with Eqs.~(\ref{eq:34}) and (\ref{eq:35}), one finds $U_1=\lambda U_0$, and hence
\begin{equation}
U_n=\lambda^nU_0,
\qquad n\geq1,
\qquad |\lambda|<1.
\end{equation}
Therefore,
\begin{equation}
\frac{U_n}{V}
=
\lambda^n
\frac{k_e}{k_e+D_s(\omega)}
\longrightarrow0
\end{equation}
as $|D_s(\omega)|/k_e\rightarrow\infty$. Suppression of the boundary-atom motion is therefore accompanied by suppression of motion throughout the semi-infinite backbone. In the limiting case, this corresponds to complete dynamical suppression of vibration transfer from the edge resonator into the backbone.

\subsection{Generalization to a periodic array of edge resonators}

\begin{figure}[htbp]
    \centering
    \includegraphics[width=0.62\textwidth]{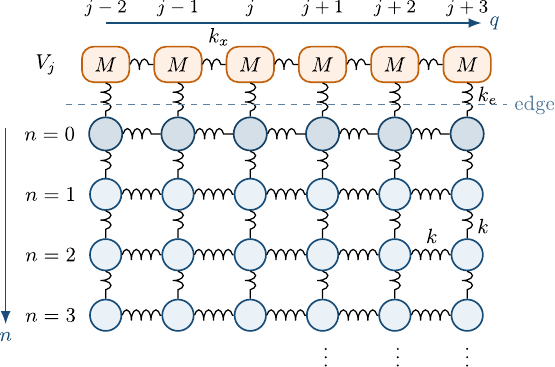}
    \caption{A periodic array of edge resonators coupled to a semi-infinite atomic backbone. The index $j$ labels cells along the edge, $n$ labels atomic rows extending into the backbone, and $q$ is the longitudinal wave number. The stiffness $k_x$ couples neighboring edge resonators, while $k_e$ couples them to the backbone.}
    \label{fig:periodic_edge}
\end{figure}

The one-dimensional model considered above can be generalized to a system consisting of an infinite periodic row of identical semi-infinite atomic chains (see Fig.~\ref{fig:periodic_edge}). This construction includes wave propagation along the boundary and makes it possible to examine how edge localization depends on the longitudinal wave number.

Let $j$ label the unit cells along the boundary, and let $n=0,1,2,\dots$ specify the atomic position in the direction perpendicular to the boundary. Denote the displacement of the $n$th atom in the $j$th chain by $u_{n,j}(t)$ and the displacement of the edge mass $M$ attached to the boundary atom by $V_j(t)$.

As before, neighboring atoms in the direction into the backbone are connected by springs of stiffness $k$, and the edge mass is connected to the atom $u_{0,j}$ by a spring of stiffness $k_e$. In addition, neighboring atomic chains are assumed to be connected along the boundary by springs of the same stiffness $k$. A longitudinal stiffness $k_x$ is introduced to describe the interaction between neighboring edge masses. The stiffness of the local foundation is set to zero, $K=0$.

For the interior backbone atoms ($n\geq1$), the equations of motion are
\begin{equation}
m\ddot{u}_{n,j}
=
-k\left(2u_{n,j}-u_{n-1,j}-u_{n+1,j}\right)
-k\left(2u_{n,j}-u_{n,j-1}-u_{n,j+1}\right).
\end{equation}
For the boundary atom $n=0$, we obtain
\begin{equation}
m\ddot{u}_{0,j}
=
-k\left(u_{0,j}-u_{1,j}\right)
-k\left(2u_{0,j}-u_{0,j-1}-u_{0,j+1}\right)
-k_e\left(u_{0,j}-V_j\right).
\end{equation}
Finally, the motion of the edge mass is governed by
\begin{equation}
M\ddot{V}_j
=
-k_e\left(V_j-u_{0,j}\right)
-k_x\left(2V_j-V_{j-1}-V_{j+1}\right).
\end{equation}

\subsection{Bloch representation}

Because the system is translationally invariant along the boundary, its solutions can be sought as Bloch waves:
\begin{equation}
u_{n,j}(t)
=
U_n e^{i(qj-\omega t)},
\qquad
V_j(t)
=
V e^{i(qj-\omega t)},
\end{equation}
where $q$ is the dimensionless wave number along the boundary. For neighboring cells,
\begin{equation}
2u_{n,j}-u_{n,j-1}-u_{n,j+1}
=
4\sin^2\left(\frac{q}{2}\right)u_{n,j}.
\end{equation}
It is therefore convenient to introduce the longitudinal dynamic stiffnesses of the atomic backbone and the edge subsystem:
\begin{equation}
\kappa(q)
=
4k\sin^2\left(\frac{q}{2}\right),
\qquad
\kappa_x(q)
=
4k_x\sin^2\left(\frac{q}{2}\right).
\label{eq:kappa_def}
\end{equation}
In the Bloch representation, the equation of motion of the edge mass is
\begin{equation}
\left[
k_e+\kappa_x(q)-M\omega^2
\right]V
-k_eU_0
=
0.
\end{equation}
It follows directly that
\begin{equation}
U_0
=
\frac{k_e+\kappa_x(q)-M\omega^2}{k_e}\,V.
\label{eq:U0_periodic}
\end{equation}
Introduce the $q$-dependent eigenfrequency of the edge subsystem,
\begin{equation}
\omega_A^2(q)
=
\frac{k_e+\kappa_x(q)}{M}
=
\frac{k_e+4k_x\sin^2(q/2)}{M}.
\label{eq:omegaA}
\end{equation}
Equation~\eqref{eq:U0_periodic} can then be written as
\begin{equation}
U_0
=
\frac{k_e+\kappa_x(q)}{k_e}
\left[
1-\frac{\omega^2}{\omega_A^2(q)}
\right]V.
\label{eq:U0_periodic_resonance}
\end{equation}
This relation directly generalizes the result obtained above for a single edge resonator. Indeed, in the absence of longitudinal coupling between the edge masses, $k_x=0$, we have
\begin{equation}
\kappa_x(q)=0,
\qquad
\omega_A(q)=\omega_0,
\qquad
\omega_0^2=\frac{k_e}{M}.
\end{equation}
In this case, Eq.~\eqref{eq:U0_periodic_resonance} reduces exactly to the relation for a single edge resonator:
\begin{equation}
U_0
=
\left(
1-\frac{\omega^2}{\omega_0^2}
\right)V.
\label{eq:U0_single_limit}
\end{equation}

Thus, for $k_x=0$, the local dynamics of each edge resonator is identical to that of an isolated resonator: its resonance frequency $\omega_0$ is independent of $q$. For $k_x\neq0$, longitudinal coupling between the edge masses turns this single resonance frequency into a dispersive edge branch $\omega_A(q)$. In both cases, the boundary-atom amplitude vanishes under the same resonance condition,
\begin{equation}
\omega=\omega_A(q),
\end{equation}
which reduces to $\omega=\omega_0$ when $k_x=0$.

\subsection{Five-layer ribbon model with edge resonators}

Consider a ribbon of finite width consisting of five identical atomic layers with atomic mass $m$. The ribbon is infinite and periodic in the longitudinal direction, while the transverse layers are labeled by
\begin{equation}
n=-2,-1,0,1,2.
\end{equation}
Identical edge masses $M$ are attached to the two outermost layers, $n=\pm2$, by springs of stiffness $k_e$. Neighboring atomic layers are coupled by springs of stiffness $k$. Longitudinal coupling between the edge masses is absent, $k_x=0$, and the local foundation stiffness is set to zero, $K=0$.

Because the ribbon is translationally invariant along its length, we seek Bloch-wave solutions,
\begin{equation}
u_{n,j}(t)=U_n e^{i(qj-\omega t)},
\end{equation}
and, for the edge masses,
\begin{equation}
V_j^{\pm}(t)=V_{\pm}e^{i(qj-\omega t)},
\end{equation}
where $q$ is the dimensionless wave number along the ribbon. The longitudinal dynamic stiffness of the atomic backbone, $\kappa(q)$, is defined by Eq.~\eqref{eq:kappa_def}.

Introduce the vector of complex amplitudes
\begin{equation}
\mathbf{U}
=
\begin{pmatrix}
V_- &
U_{-2} &
U_{-1} &
U_0 &
U_1 &
U_2 &
V_+
\end{pmatrix}^{T}.
\end{equation}

\begin{figure}[htbp]
    \centering
    \includegraphics[width=0.62\textwidth]{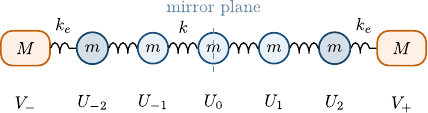}
    \caption{Cross section of a five-layer ribbon with two edge resonators. The dashed line marks the mirror-symmetry plane passing through the central layer $U_0$.}
    \label{fig:finite_strip}
\end{figure}

The frequency-domain equations of motion are
\begin{equation}
\mathbf{D}(q,\omega)\mathbf{U}=0,
\label{eq:full_system}
\end{equation}
where
\begin{equation}
\mathbf{D}(q,\omega)
=
\begin{pmatrix}
d_M & -k_e & 0 & 0 & 0 & 0 & 0\\
-k_e & d_b & -k & 0 & 0 & 0 & 0\\
0 & -k & d & -k & 0 & 0 & 0\\
0 & 0 & -k & d & -k & 0 & 0\\
0 & 0 & 0 & -k & d & -k & 0\\
0 & 0 & 0 & 0 & -k & d_b & -k_e\\
0 & 0 & 0 & 0 & 0 & -k_e & d_M
\end{pmatrix},
\end{equation}
and
\begin{equation}
d_M(\omega)=k_e-M\omega^2,
\qquad
d_b(q,\omega)=k+k_e+\kappa(q)-m\omega^2,
\qquad
d(q,\omega)=2k+\kappa(q)-m\omega^2.
\label{eq:diagonal_elements}
\end{equation}
The full spectrum of the system is determined by
\begin{equation}
\det\mathbf{D}(q,\omega)=0.
\label{eq:full_secular}
\end{equation}

\subsubsection{Lower acoustic branch}

Because of mirror symmetry about the central layer, the eigenmodes separate into symmetric and antisymmetric sectors. The lower acoustic branch belongs to the symmetric sector, for which
\begin{equation}
V_-=V_+=V,
\qquad
U_{-2}=U_2,
\qquad
U_{-1}=U_1.
\end{equation}
It is therefore sufficient to consider four independent amplitudes,
\begin{equation}
\mathbf{U}_+
=
\begin{pmatrix}
V & U_2 & U_1 & U_0
\end{pmatrix}^{T}.
\end{equation}
The corresponding dynamical matrix is
\begin{equation}
\mathbf{D}_+(q,\omega)
=
\begin{pmatrix}
k_e-M\omega^2 & -k_e & 0 & 0\\
-k_e & k+k_e+\kappa(q)-m\omega^2 & -k & 0\\
0 & -k & d(q,\omega) & -k\\
0 & 0 & -2k & d(q,\omega)
\end{pmatrix}.
\label{eq:symmetric_matrix}
\end{equation}

The internal amplitudes $U_1$ and $U_0$ can be eliminated analytically. The last two equations give
\begin{equation}
U_1=
\frac{k\,d(q,\omega)}
{d^2(q,\omega)-2k^2}\,U_2,
\qquad
U_0=
\frac{2k^2}
{d^2(q,\omega)-2k^2}\,U_2.
\label{eq:internal_amplitudes}
\end{equation}
After substitution, the effect of the internal layers on the outermost layer $U_2$ is described by the effective dynamic stiffness
\begin{equation}
D_s(q,\omega)
=
k+\kappa(q)-m\omega^2
-
\frac{k^2d(q,\omega)}
{d^2(q,\omega)-2k^2}.
\label{eq:Ds}
\end{equation}
The symmetric problem is therefore reduced exactly to
\begin{equation}
\begin{pmatrix}
k_e-M\omega^2 & -k_e\\
-k_e & k_e+D_s(q,\omega)
\end{pmatrix}
\begin{pmatrix}
V\\
U_2
\end{pmatrix}
=0.
\label{eq:reduced_system}
\end{equation}
The condition for a nontrivial solution gives 
\begin{equation}
\left(k_e-M\omega^2\right)
\left[k_e+D_s(q,\omega)\right]
-k_e^2=0.
\end{equation}
Introduce the characteristic frequency of the edge resonator,
\begin{equation}
\omega_0^2=\frac{k_e}{M}.
\label{eq:omega0}
\end{equation}
The exact dispersion relation for the symmetric sector can then be written in the compact form
\begin{equation}
\omega^2
=
\omega_0^2
\frac{D_s(q,\omega)}
{k_e+D_s(q,\omega)}.
\label{eq:exact_dispersion}
\end{equation}

\subsubsection{Approximate solution of the dispersion equation}

To follow explicitly how the lower acoustic branch changes as the wave number increases, consider an approximation for $D_s(q,\omega)$. Introduce
\begin{equation}
x=\kappa(q)-m\omega^2.
\end{equation}
Then $d=2k+x$, and Eq.~\eqref{eq:Ds} becomes
\begin{equation}
D_s
=
k+x-
\frac{k^2(2k+x)}
{(2k+x)^2-2k^2}.
\end{equation}
Expansion in $x/k$, i.e., about the acoustic branch of the bare backbone, gives
\begin{equation}
D_s
=
\frac{5}{2}x
+
O\left(\frac{x^2}{k}\right),
\end{equation}
so that to first order
\begin{equation}
D_s(q,\omega)
\simeq
\frac{5}{2}
\left[\kappa(q)-m\omega^2\right].
\label{eq:Ds_approx}
\end{equation}

The accuracy of this approximation is controlled by the small parameter $x/k$. In the crossover region leading to the edge-resonant regime, its characteristic value can be estimated as
\begin{equation}
\frac{x_c}{k}
\sim
\frac{2k_e}{Nk},
\end{equation}
where $N$ is the number of backbone rows in the ribbon ($N=5$ in the present case). The approximation therefore becomes more accurate as the ratio $k_e/k$ decreases, that is, as the edge-resonator coupling becomes weaker relative to the stiffness of the atomic backbone. At fixed $k_e$ and $k$, its accuracy also improves as the ribbon width $N$ increases.

Substitution of Eq.~\eqref{eq:Ds_approx} into Eq.~\eqref{eq:exact_dispersion} gives the biquadratic equation
\begin{equation}
Mm\omega^4
-
\left[
M\kappa(q)+k_em+\frac{2}{5}Mk_e
\right]\omega^2
+
k_e\kappa(q)
=0.
\label{eq:approx_dispersion}
\end{equation}
This equation directly displays the crossover of the lower branch from the acoustic regime to the resonant plateau.

In the long-wavelength limit $q\rightarrow0$,
\begin{equation}
\kappa(q)\simeq kq^2,
\end{equation}
and the lower root of Eq.~\eqref{eq:approx_dispersion} is
\begin{equation}
\omega_-^2(q)
\simeq
\frac{5\kappa(q)}{5m+2M}.
\end{equation}
Consequently,
\begin{equation}
\omega_-(q)
\simeq
q\sqrt{\frac{5k}{5m+2M}},
\qquad q\rightarrow0.
\label{eq:acoustic_limit}
\end{equation}
Thus, at small $q$, the lower branch is an ordinary acoustic mode with linear dispersion.

As $q$ increases, the dynamic stiffness of the backbone, $\kappa(q)$, also increases. When the term $M\kappa(q)$ becomes dominant in the coefficient of $\omega^2$, that is, when
\begin{equation}
M\kappa(q)
\gg
k_em+\frac{2}{5}Mk_e,
\label{eq:large_q_condition}
\end{equation}
the leading part of Eq.~\eqref{eq:approx_dispersion} becomes
\begin{equation}
-M\kappa(q)\omega^2
+
k_e\kappa(q)
\simeq0.
\end{equation}
After cancellation of $\kappa(q)$, this gives
\begin{equation}
\omega_-^2(q)\simeq\frac{k_e}{M}
=\omega_0^2.
\label{eq:plateau}
\end{equation}
Hence,
\begin{equation}
\omega_-(q)\simeq\omega_0.
\end{equation}

The approximate dispersion equation therefore describes two characteristic regimes of the lower branch:
\begin{equation}
\omega_-(q)
\simeq
\begin{cases}
\displaystyle
q\sqrt{\dfrac{5k}{5m+2M}},
& q\ll q_c,\\[3mm]
\displaystyle
\sqrt{\dfrac{k_e}{M}}=\omega_0,
& q\gg q_c.
\end{cases}
\label{eq:two_limits}
\end{equation}
At small wave numbers, the branch has the usual acoustic character, $\omega_-\propto q$. As $q$ increases, its frequency saturates and approaches the resonance frequency $\omega_0$ of the edge complex.

The characteristic crossover wave vector $q_c$ can be estimated by equating the two contributions to the coefficient of $\omega^2$ in Eq.~\eqref{eq:approx_dispersion}:
\begin{equation}
M\kappa(q_c)
\sim
k_em+\frac{2}{5}Mk_e.
\end{equation}
Using
\begin{equation}
\kappa(q)=4k\sin^2\left(\frac{q}{2}\right),
\end{equation}
we obtain
\begin{equation}
\sin^2\left(\frac{q_c}{2}\right)
\sim
\frac{k_e}{4k}
\left(
\frac{m}{M}+\frac{2}{5}
\right),
\end{equation}
and hence
\begin{equation}
q_c
\sim
2\arcsin
\sqrt{
\frac{k_e}{4k}
\left(
\frac{m}{M}+\frac{2}{5}
\right)
}.
\label{eq:qc}
\end{equation}

Thus, $q_c$ specifies the characteristic wave-number scale near which the lower acoustic branch crosses over from linear growth to the resonant plateau. Because this crossover is smooth, $q_c$ should be regarded as an estimate of the center of the hybridization region rather than as a strict critical point.

\subsubsection{Spatial localization}

The change in the shape of the dispersion branch is accompanied by a change in its eigenvector. The first equation of the exact reduced system, Eq.~\eqref{eq:reduced_system}, is
\begin{equation}
\left(k_e-M\omega^2\right)V-k_eU_2=0.
\end{equation}
Using the definition of $\omega_0$, this gives directly
\begin{equation}
\frac{U_2}{V}
=
1-\frac{\omega^2}{\omega_0^2}.
\label{eq:edge_ratio}
\end{equation}

In the long-wavelength acoustic regime, $\omega\ll\omega_0$, and therefore
\begin{equation}
\frac{U_2}{V}\simeq1.
\end{equation}
The edge masses and the outermost atomic layers move almost together, and the oscillation is distributed across the full width of the ribbon.

As the lower branch approaches the resonant plateau, Eq.~\eqref{eq:plateau} gives
\begin{equation}
\omega_-(q)\rightarrow\omega_0,
\end{equation}
and hence
\begin{equation}
\frac{U_2}{V}\rightarrow0.
\end{equation}
The motion of the outermost atomic layer is therefore suppressed relative to the motion of the edge mass. According to Eq.~\eqref{eq:internal_amplitudes}, the amplitudes of the internal layers are proportional to $U_2$ and decrease as well. In the limit of strong edge localization, the eigenvector becomes
\begin{equation}
\left(
V,U_2,U_1,U_0,U_1,U_2,V
\right)
\longrightarrow
\left(
V,0,0,0,0,0,V
\right).
\label{eq:localized_vector}
\end{equation}

The dispersion and the spatial character of the lower branch therefore change simultaneously. At small $q$, it is an acoustic mode distributed across the ribbon, with $\omega\propto q$. Within the crossover region, amplitude is redistributed between the atomic backbone and the edge degrees of freedom. For $q>q_c$, the same branch becomes predominantly localized at the edge and its frequency approaches $\omega_0$. Thus, the crossover from a linear acoustic dispersion to an almost flat edge branch represents the hybridization of the collective acoustic mode of the ribbon with a local resonance of the edge complex.

\subsubsection{Width dependence of the crossover to local edge vibrations}

The result generalizes naturally to a ribbon of arbitrary width. Let the atomic backbone contain $N$ identical layers, with identical resonators of mass $M$ and coupling stiffness $k_e$ attached to the two outermost layers. In the long-wavelength limit, the effective dynamic stiffness of the backbone is
\begin{equation}
D_s(q,\omega)
\simeq
\frac{N}{2}
\left[
\kappa(q)-m\omega^2
\right].
\end{equation}
For $N=5$, this expression reduces to the result derived above, $D_s\simeq(5/2)(\kappa-m\omega^2)$.

Substitution into the dispersion equation gives
\begin{equation}
Mm\omega^4
-
\left[
M\kappa(q)+k_em+\frac{2}{N}Mk_e
\right]\omega^2
+
k_e\kappa(q)
=0.
\end{equation}
At small $q$, the lower root remains acoustic:
\begin{equation}
\omega_-^2(q)
\simeq
\frac{N\kappa(q)}{Nm+2M},
\end{equation}
whereas at large $q$ it approaches the edge-resonator frequency,
\begin{equation}
\omega_-(q)\simeq
\omega_0
=
\sqrt{\frac{k_e}{M}}.
\end{equation}
The characteristic crossover wave vector is determined by
\begin{equation}
\kappa(q_c)
\sim
k_e
\left(
\frac{m}{M}+\frac{2}{N}
\right).
\end{equation}
Since
\begin{equation}
\kappa(q)=4k\sin^2\left(\frac{q}{2}\right),
\end{equation}
we obtain
\begin{equation}
q_c(N)
\sim
2\arcsin
\sqrt{
\frac{k_e}{4k}
\left(
\frac{m}{M}+\frac{2}{N}
\right)
}.
\label{eq:qc_width}
\end{equation}

Thus, as the ribbon becomes wider, $q_c$ decreases: the crossover from a collective acoustic mode to a predominantly edge-localized vibration occurs at a smaller wave number. This width dependence gradually saturates. In the limit $N\rightarrow\infty$,
\begin{equation}
q_c(\infty)
\sim
2\arcsin
\sqrt{
\frac{k_em}{4kM}
}.
\end{equation}
Increasing the ribbon width therefore does not eliminate edge localization, it merely shifts the crossover region toward a finite limiting wave number.

\subsubsection{Critical influence of longitudinal coupling between edge resonators}

The analysis of the five-layer model and its generalization to a ribbon of width $N$ has so far assumed that neighboring edge resonators are not coupled, i.e., $k_x=0$. We now consider $k_x\neq0$. Longitudinal coupling gives the edge subsystem an additional $q$-dependent dynamic stiffness,
\begin{equation}
\kappa_x(q)
=
4k_x\sin^2\left(\frac{q}{2}\right)
=
\frac{k_x}{k}\,\kappa(q).
\end{equation}
The eigenfrequency of the isolated edge subsystem therefore becomes dispersive:
\begin{equation}
\omega_A^2(q)
=
\frac{k_e+\kappa_x(q)}{M}.
\label{eq:omegaA_kx}
\end{equation}

Let $U_b$ denote the amplitude of the outermost atomic layer of the backbone directly coupled to the edge resonator ($U_b=U_2$ for the five-layer model considered above). The reduced system is
\begin{equation}
\begin{pmatrix}
k_e+\kappa_x(q)-M\omega^2 & -k_e\\
-k_e & k_e+D_s(q,\omega)
\end{pmatrix}
\begin{pmatrix}
V\\
U_b
\end{pmatrix}
=0.
\end{equation}
The second equation gives directly
\begin{equation}
\frac{U_b}{V}
=
\frac{k_e}{k_e+D_s(q,\omega)}.
\label{eq:edge_ratio_kx}
\end{equation}
Strong edge localization therefore requires
\begin{equation}
D_s(q,\omega)\gg k_e,
\end{equation}
because in this case $|U_b/V|\ll1$: motion of the atomic backbone is suppressed relative to that of the edge resonator.

When $|U_b/V|\ll1$, the first equation of the reduced system gives
\begin{equation}
k_e+\kappa_x(q)-M\omega^2\simeq0,
\end{equation}
and the localized branch therefore approaches the dispersion of the edge subsystem:
\begin{equation}
\omega^2(q)\simeq\omega_A^2(q)
=
\frac{k_e+\kappa_x(q)}{M}.
\label{eq:localized_edge_dispersion}
\end{equation}
For $k_x=0$, this expression reduces to the flat resonant plateau $\omega\simeq\omega_0=\sqrt{k_e/M}$. For $k_x\neq0$, the localized branch is generally no longer strictly flat and inherits the weak dispersion of the edge subsystem.

We now determine when the edge localization, once formed, persists as $q$ increases further. For a ribbon of width $N$, use the approximation derived above,
\begin{equation}
D_s(q,\omega)
\simeq
\frac{N}{2}
\left[
\kappa(q)-m\omega^2
\right].
\end{equation}
Within the established edge-mode regime, Eq.~\eqref{eq:localized_edge_dispersion} may be used. Then
\begin{align}
D_s(q,\omega)
&\simeq
\frac{N}{2}
\left[
\kappa(q)
-\frac{m}{M}
\left(
k_e+\kappa_x(q)
\right)
\right]
\\
&=
\frac{N}{2}
\left[
\left(
1-\frac{m k_x}{M k}
\right)\kappa(q)
-\frac{m}{M}k_e
\right].
\label{eq:Ds_kx}
\end{align}

This expression shows directly the role of the longitudinal stiffness $k_x$. If
\begin{equation}
1-\frac{m k_x}{M k}>0,
\qquad\text{that is,}\qquad
\frac{k_x}{k}<\frac{M}{m},
\label{eq:kx_condition}
\end{equation}
then $\kappa(q)$ increases with $q$, and so does $D_s$. According to Eq.~\eqref{eq:edge_ratio_kx}, the ratio $|U_b/V|$ continues to decrease. Thus, once the branch has crossed into the edge regime, localization does not disappear as the wave number increases further.

The physical meaning of this condition can be seen directly by comparing the dispersion of the edge-resonator subsystem with that of the atomic backbone. The edge-resonance frequency varies with wave number as
\begin{equation}
\omega_A^2(q)
=
\omega_0^2+
\frac{4k_x}{M}
\sin^2\left(\frac{q}{2}\right),
\end{equation}
whereas the characteristic $q$ dependence of the acoustic backbone branches is set by
\begin{equation}
\frac{\kappa(q)}{m}
=
\frac{4k}{m}
\sin^2\left(\frac{q}{2}\right).
\end{equation}

When
\begin{equation}
\frac{k_x}{M}\ll\frac{k}{m},
\end{equation}
the edge-resonance branch disperses much more weakly than the backbone branches. As $q$ increases, an acoustic branch can therefore enter the frequency range of the edge resonance, after which its eigenvector becomes predominantly edge-like.

As $k_x$ increases, $\omega_A(q)$ itself shifts upward more rapidly. Consequently, the edge-resonance region is reached at a larger $q$, which is reflected directly in the increase of $q_c(N,k_x)$. If the longitudinal coupling between the edge resonators is sufficiently strong, the resonance branch may leave the frequency range of the acoustic branch under consideration before pronounced edge localization develops. In that case, the corresponding $q_c$ lies outside the first Brillouin zone.

Equation~\eqref{eq:Ds_kx} also yields an estimate of the characteristic crossover wave number. Strong localization requires $D_s\gg k_e$, leading to
\begin{equation}
\left(
1-\frac{m k_x}{M k}
\right)\kappa(q)
\gg
k_e
\left(
\frac{m}{M}+\frac{2}{N}
\right).
\end{equation}
We estimate $q_c$ by replacing this inequality with an equality of the two scales:
\begin{equation}
\left(
1-\frac{m k_x}{M k}
\right)\kappa(q_c)
\sim
k_e
\left(
\frac{m}{M}+\frac{2}{N}
\right).
\end{equation}
Using $\kappa(q)=4k\sin^2(q/2)$, we obtain
\begin{equation}
q_c(N,k_x)
\sim
2\arcsin
\sqrt{
\frac{k_e}{4k}
\,
\frac{
\dfrac{m}{M}+\dfrac{2}{N}
}{
1-\dfrac{m k_x}{M k}
}
}.
\label{eq:qc_kx}
\end{equation}

For $k_x=0$, Eq.~\eqref{eq:qc_kx} reduces exactly to the estimate obtained above,
\begin{equation}
q_c(N)
\sim
2\arcsin
\sqrt{
\frac{k_e}{4k}
\left(
\frac{m}{M}+\frac{2}{N}
\right)
}.
\end{equation}
Increasing $k_x$ therefore reduces the dynamical contrast between the atomic backbone and the edge subsystem and shifts the crossover to larger wave numbers. In the limit $mk_x/(Mk)\rightarrow1$, the characteristic value of $q_c$ grows rapidly and may move beyond the first Brillouin zone.

In addition, the edge-resonance frequency must lie within the frequency range of the corresponding phonon branch.

For this crossover to occur within the first Brillouin zone, the expression under the square root in Eq.~\eqref{eq:qc_kx} must not exceed unity:
\begin{equation}
\frac{k_e}{4k}
\,
\frac{
\dfrac{m}{M}+\dfrac{2}{N}
}{
1-\dfrac{m k_x}{M k}
}
\lesssim1.
\label{eq:qc_existence}
\end{equation}
If this condition is violated, the characteristic crossover scale lies outside the accessible interval $0\leq q\leq\pi$, and the fully developed edge regime does not form within this approximation.

Thus, the essential condition for the mechanism considered here is not the strict equality $k_x=0$, but rather a dynamical contrast between the backbone and the edge subsystem. The smaller the ratio $k_x/k$, the weaker the dispersion of the edge resonator's eigenfrequency, the earlier the crossover to the edge regime occurs, and the more strongly the motion of the atomic backbone is suppressed beyond it. The case $k_x=0$ is the limiting case of maximum dynamical contrast, in which the localized edge branch becomes almost flat.

\subsection{Resonant limit and edge localization}

Unlike a driven frequency, which can be set exactly equal to $\omega_0$,
an eigenfrequency must satisfy the dispersion equation
\begin{equation}
\left(k_e-M\omega^2\right)
\left[k_e+D_s(q,\omega)\right]
-k_e^2=0.
\end{equation}
Using
\begin{equation}
\omega_0^2=\frac{k_e}{M},
\end{equation}
we obtain
\begin{equation}
\frac{\omega^2}{\omega_0^2}
=
\frac{D_s(q,\omega)}
{k_e+D_s(q,\omega)}.
\label{eq:omega_Ds_relation}
\end{equation}
Consequently,
\begin{equation}
1-\frac{\omega^2}{\omega_0^2}
=
\frac{k_e}{k_e+D_s(q,\omega)}.
\label{eq:detuning_Ds}
\end{equation}
For finite $D_s$, the eigenfrequency does not reach $\omega_0$ exactly. However, when $D_s\gg k_e$, the detuning from the resonance frequency becomes small and the eigenbranch approaches the resonant plateau.

The physical meaning of this limit follows directly from the first equation of the reduced system:
\begin{equation}
\frac{U_b}{V}
=
1-\frac{\omega^2}{\omega_0^2},
\end{equation}
where $U_b$ is the amplitude of the outermost atomic layer of the backbone ($U_b=U_2$ for the five-layer model). Together with Eq.~\eqref{eq:detuning_Ds}, this gives the exact relation
\begin{equation}
\frac{U_b}{V}
=
\frac{k_e}{k_e+D_s(q,\omega)}.
\label{eq:UbV_Ds}
\end{equation}

Thus, an increase in the effective dynamic stiffness of the backbone simultaneously brings the eigenfrequency closer to $\omega_0$ and suppresses the backbone motion:
\begin{equation}
D_s\gg k_e
\qquad\Longrightarrow\qquad
\omega\simeq\omega_0,
\qquad
\left|\frac{U_b}{V}\right|\ll1.
\end{equation}
For the five-layer model, the internal amplitudes $U_1$ and $U_0$ are proportional to $U_2$. Therefore, suppressing $U_2$ reduces the amplitudes throughout the atomic backbone, while the motion of the edge resonators remains finite.

Edge localization therefore does not require the frequency to reach $\omega_0$ exactly. Once the eigenfrequency is sufficiently close to the local resonance, the ratio $|U_b/V|$ becomes small and the weight of the edge resonators in the eigenvector rapidly approaches unity. In other words, the approach of the eigenbranch to the local edge-resonance frequency and the transfer of oscillation amplitude from the atomic backbone to the edge are two aspects of the same dynamical process.

\section{Atomistic calculations and electronic response}
\label{sec:methods}
\subsection{Reference geometry and interatomic-potential selection}
The equilibrium lattice parameters of the two-dimensional biphenylene network were used to compare candidate machine-learned interatomic potentials with density-functional theory (DFT). The reference structure was optimized using the Vienna \textit{Ab initio} Simulation Package (VASP)~\cite{Kresse1996VASP} with the B3LYP hybrid functional. The core--valence interaction was described using projector augmented-wave (PAW) potentials~\cite{Kresse1999PAW}. The plane-wave kinetic-energy cutoff was $E_{\mathrm{cut}}=600$ eV (\texttt{ENCUT}=600), the Brillouin zone was sampled with a $\Gamma$-centered $5\times4\times1$ mesh, and the vacuum separation was approximately $20$~\AA. The maximum residual force was reduced below $5\times10^{-3}$ eV/\AA.

The tested models were PET-MAD-S, PET-OAM-XL, PET-OMAD-S, and PET-SPICE-L from UPET~\cite{Bigi2026}, and the FAIR Chemistry UMA-M model for the OMC and OMol tasks~\cite{Wood2025}. Each model was used to optimize the same reference structure. PET-MAD-S gave the closest agreement with the DFT lattice parameters and was selected for the ribbon calculations.

\subsection{Freestanding and hBN-supported ribbons}
Ribbon width is specified by the number $N_h$ of hexagons across the ribbon, as illustrated in Fig.~1 of the Letter. Periodic boundary conditions are imposed along the ribbon axis. For the supported system, a nearly commensurate ribbon--hBN supercell contains approximately $2\times10^3$ atoms and satisfies
\begin{equation}
\frac{|L_{\mathrm{BPH}}-L_{\mathrm{hBN}}|}{L_{\mathrm{BPH}}}\leq 0.003.
\label{eq:mismatch}
\end{equation}
The transverse substrate dimension separates neighboring periodic images of the ribbon. The initial ribbon--substrate separation is $4$~\AA. The hBN coordinates are fixed during structural relaxation, while the ribbon coordinates are optimized.

Phonon frequencies and polarization vectors are obtained with Phonopy~\cite{phonopy2015,phonopy2023,phonopy20232}. The electronic spectra of displaced configurations are calculated using density-functional tight binding (DFTB)~\cite{Hourahine2020} with PTBP Slater--Koster parameters~\cite{Cui2024PTBP}, without self-consistent charge iterations. For supported ribbons, only ribbon atoms are retained in the electronic calculation. Thus, the substrate affects the calculated gap through the ribbon geometry and vibrational model.

\subsection{Frozen displacements and quantum normalization}

For each normal mode $i$, we define the quantum-normalized displacement pattern following Refs.~\cite{Capaz2005,Cohen1986}:
\begin{equation}
x_i(\alpha,\kappa)=
\left(\frac{\hbar}{M_\kappa\omega_i}\right)^{1/2}
\epsilon_i(\alpha,\kappa),
\qquad
\sum_{\alpha,\kappa}
\left|\epsilon_i(\alpha,\kappa)\right|^2=1,
\label{eq:normalization}
\end{equation}
where $\alpha$ denotes a Cartesian component, $\kappa$ labels an atom of mass $M_\kappa$, and $\epsilon_i(\alpha,\kappa)$ is the corresponding component of the phonon polarization vector. The index $i$ runs over the complete set of normal modes of the periodic supercell. The vector $\mathbf{x}_i$ is the atom-resolved displacement pattern corresponding to the quantum-normalized normal coordinate $Q_i$ used in the Letter.

For each mode, two oppositely distorted configurations are constructed according to
\begin{equation}
R_{\alpha\kappa}^{(\pm i)}
=
R_{\alpha\kappa}^{(0)}
\pm c_i x_i(\alpha,\kappa).
\label{eq:distorted-structures}
\end{equation}
The dimensionless coefficient $c_i$ is selected separately for each mode so that the maximum change in the distance between neighboring atoms is $0.0125$~\AA. Within the quadratic approximation, the mode-resolved gap-renormalization coefficient is then obtained as
\begin{equation}
W_i=
\frac{
E_g(+c_i\mathbf{x}_i)+E_g(-c_i\mathbf{x}_i)
-2E_g^{\mathrm{stat}}
}{2c_i^2}.
\label{eq:Wi}
\end{equation}
The symmetric combination eliminates all contributions that are odd in the phonon coordinate, while division by $c_i^2$ removes the mode-dependent numerical scaling of the frozen displacement. Thus, $W_i$ is independent of the particular value of $c_i$ within the quadratic regime and represents the change in the band-edge separation per phonon occupation.

The thermal average separates into static, zero-point, and additional thermal contributions:
\begin{align}
E_g(T)&=E_g^{\mathrm{stat}}
+\sum_iW_i\left[n_i(T)+\frac12\right],
\label{eq:thermal}\\
n_i(T)&=
\frac{1}{\exp(\hbar\omega_i/k_{\mathrm B}T)-1},
\\
\Delta E_g^{\mathrm{ZP}}&=\frac12\sum_iW_i,
\qquad
\Delta E_g(T)=E_g(T)-E_g(0)
=\sum_iW_i n_i(T).
\label{eq:ZP-thermal}
\end{align}
Zero-point renormalization samples the complete set of mode weights, whereas the additional thermal shift weights the same modes by their Bose occupations and therefore emphasizes low-frequency vibrations at low temperature. Consequently, a mode that dominates the initial thermal variation need not dominate the total zero-point correction.

\section{H--H edge termination: a control calculation}
\begin{figure}
\centering
\includegraphics[width=0.92\textwidth]{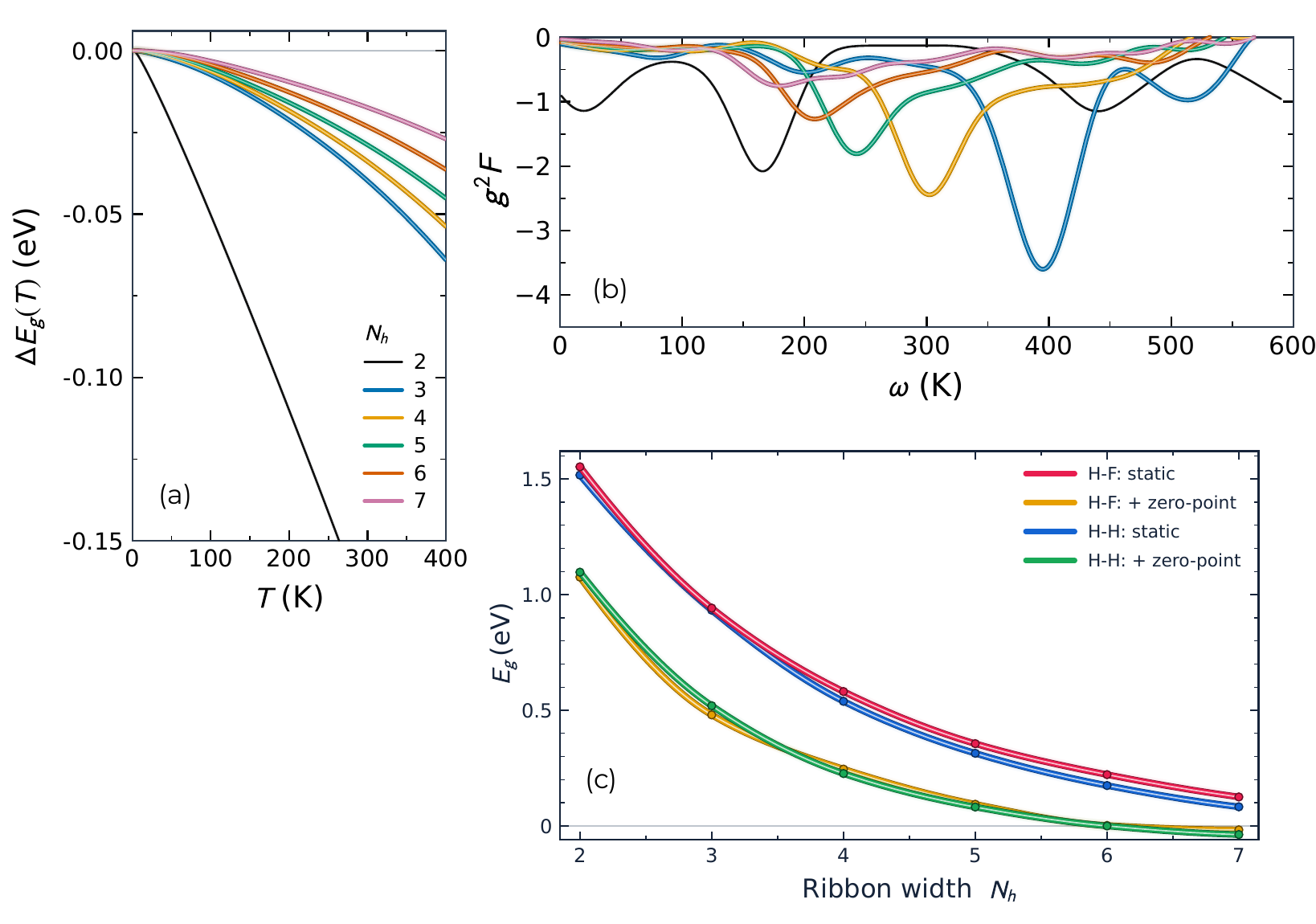}
\caption{Control calculations for freestanding H--H-terminated ribbons.
(a) Temperature-dependent change in the band-edge separation.
(b) Signed gap-response spectrum $g^2F$. The delta functions are replaced by Gaussians with $\sigma=23$~K, as in the Letter.
(c) Static and zero-point-renormalized band-edge separations
for freestanding H--H- and H--F-terminated ribbons.}
\label{fig:hh_control}
\end{figure}

As a control calculation, we examined freestanding biphenylene ribbons
with H--H edge termination (Fig.~\ref{fig:hh_control}). Their thermal
gap narrowing persists, but is generally weaker than in freestanding
H--F ribbons [Fig.~\ref{fig:hh_control}(a)]. Its magnitude is close to
that found for H--F ribbons on hBN in the Letter, although the H--H
ribbons considered here are freestanding.

The difference is also visible in $g^2F$ [Fig.~\ref{fig:hh_control}(b)].
With H--H termination, the pronounced low-frequency feature A
disappears and the principal spectral peaks become weaker. Feature B
remains: as in H--F ribbons, it is associated with the U-shaped segment
of the width-breathing phonon branch. Thus, changing the edge complex
weakens the thermal response while retaining the contribution from
transverse breathing of the ribbon.

At the same time, the static band-edge separations and their total
zero-point corrections remain similar for the two edge terminations
[Fig.~\ref{fig:hh_control}(c)]. The loss of feature A is barely apparent
in the zero-point correction, which includes contributions from the
entire phonon spectrum. The control calculation therefore identifies
the H--F edge complex with the pronounced feature A and the stronger
thermal response, while feature B and a substantial zero-point
correction persist with H--H termination.

\clearpage
\bibliographystyle{apsrev4-2}
\bibliography{references_supp.bib}